\documentclass{aa} 

\usepackage{graphicx}
\usepackage{bm}
\usepackage{multicol}
\graphicspath{{figs/}}

\usepackage{txfonts}
\usepackage[colorlinks,allcolors=blue]{hyperref}

\begin{document}

   \title{Origin of eclipsing time variations:
   Contributions of different modes of the
   dynamo-generated magnetic field}
   \titlerunning{Contribution of magnetic field modes to ETVs}

   \author{Felipe H. Navarrete\inst{1,2},
          Petri J. K\"apyl\"a\inst{3,2},
          Dominik R.G. Schleicher\inst{4},
          Carolina A. Ortiz\inst{4}
          \and
          Robi Banerjee\inst{1}\fnmsep
          }
    \authorrunning{Felipe H. Navarrete et al.}

   \institute{
            Hamburger Sternwarte, Universit\"at Hamburg, Gojenbergsweg 112, 21029
            Hamburg, Germany\\
            \email{felipe.navarrete@hs.uni-hamburg.de}\and
            Nordita,
            Stockholm University and KTH Royal Institute of Technology
            Hannes Alfvéns v\"ag 12, SE-106 91 Stockholm, Sweden\and
            Institut f\"ur Astrophysik, Georg-August-Universit\"at G\"ottingen,
            Friedrich-Hund-Platz 1, 37077 G\"ottingen, Germany\and
            Departamento de Astronom\'ia, Facultad de Ciencias F\'isicas y Matemáticas,
            Universidad de
            Concepci\'on, Av. Esteban Iturra s/n Barrio Universitario, Casilla 160-C, Chile
   }

   \date{Received --; accepted --}

 
  \abstract
    {The possibility to detect circumbinary planets and to
    study stellar magnetic fields through eclipsing time variations (ETVs)
    in binary stars has sparked an increase of interest in this area of
    research.}
    {We revisit the connection between
    stellar magnetic fields and the gravitational quadrupole moment $Q_{xx}$
    and compare different dynamo-generated ETV models with our simulations.}
    {We present
    magnetohydrodynamical simulations of solar mass stars with rotation periods
    of 8.3, 1.2, and 0.8 days and perform a detailed analysis of the magnetic
    and quadrupole moment using spherical harmonic decomposition.}
    {The extrema
    of $Q_{xx}$ are associated with changes in the magnetic field structure. This
    is evident in the simulation with a rotation period of 1.2 days. Its magnetic
    field has a more complex behavior than in the other models,
    as the large-scale
    nonaxisymmetric field dominates throughout the simulation and the axisymmetric
    component is predominantly hemispheric. This triggers variations in the
    density field that follow the magnetic field asymmetry with respect to the
    equator, affecting the $zz$ component of the inertia tensor, and thus
    modulating $Q_{xx}$. The magnetic fields of the two other runs are less 
    variable in time and more symmetric with respect to the equator, such that 
    the variations in the density are weaker, and
    therefore only small variations in $Q_{xx}$ are seen.}
    {If interpreted via the
    classical Applegate mechanism (tidal locking), the quadrupole moment variations
    obtained in the current simulations are about two orders of
    magnitude below those deduced from observations of
    post-common-envelope binaries. However, if no tidal locking is
    assumed, our results are
    compatible with the observed ETVs.}


   \maketitle

\section{Introduction}

Post-common-envelope binaries (PCEBs) are commonly composed of a white dwarf
and a low-mass main-sequence star. Observations of eclipses in these systems
reveal deviations from the calculated eclipsing times in approximately $90\%$
of these systems \citep{zorotovic13}, with binary period variations on the order
of $10^{-6} \ldots 10^{-7}$ modulated over periods on the order of decades.

The two main explanations, although not mutually exclusive, are the planetary
hypothesis \citep{Brinkworth06, Voelschow14} and the Applegate mechanism
\citep{Applegate92, Lanza98, Voelschow18, Lanza20}. In the planetary hypothesis,
sufficiently massive planets can force the barycenter of the binary to change
its location as they orbit, which would then explain the
observed-minus-calculated (O-C) diagram of the eclipsing times. On the other
hand, the Applegate mechanism explains the variations via the connection between
stellar magnetic fields and the gravitational quadrupole moment $Q$. The idea
behind this mechanism is that when $Q$ increases, the gravitational field also
increases. For this to happen, there must be a redistribution of angular momentum
within the star. When angular momentum is carried to the outer parts of the
convective zone (CZ), these layers rotate faster and, overall, the star
becomes more oblate, which is reflected by an increase in the gravitational
quadrupole moment. As there is no angular momentum exchange between the orbit
and the star, the orbital velocity increases and the radius decreases in order
to maintain the angular momentum of the binary.
Thus, the orbital period shortens. In order for this mechanism to work,
\cite{Applegate92} invoked the presence of a cyclic subsurface
magnetic field on
the order of 10~kG which is responsible for redistributing the
internal angular
momentum of the star.

Confirming the planetary hypothesis requires a detection of the proposed
circumbinary bodies in PCEBs either by directly imaging them, as attempted by
\cite{Hardy15}, or via indirect methods such as those employed by
\cite{Vanderbosch17}. However, these studies did not detect the proposed third
body, a brown dwarf, in V471~Tau, which is a PCEB with a Sun-like main-sequence
star and a white dwarf. It was the system \citet{Paczynski76} used to develop
the theory of PCEB formation.
Direct modeling of the Applegate mechanism is challenging, and
targeted numerical simulations that may help to understand observations have
been lacking. \cite{Navarrete20} presented the first self-consistent 3D
magnetohydrodynamical (MHD) simulations of stellar magneto-convection addressing
this problem. In that study, the time evolution
of the gravitational quadrupole moment and its correlation with the stellar
magnetic field and rotation was studied using two simulations of a solar mass
star with three and twenty times solar rotation, corresponding to rotation
periods of 8.3 and 1.2 days. However, the centrifugal force,
a key ingredient in the original Applegate mechanism, was not included in these
simulations. Nevertheless, there were still significant temporal variations
of $Q$ due to the response of the stellar structure to the dynamo-generated
magnetic field. Such variations were absent in hydrodynamical
simulations, confirming their magnetic origin.

Recently, \cite{Lanza20} presented an alternative to the Applegate mechanism
by extending the earlier work of \citet{Applegate89}.
He assumed the presence of a persistent nonaxisymmetric magnetic field inside
the CZ of the main-sequence star that was modeled as a single flux tube lying
at the equatorial plane. The density is lower within the magnetic region in
comparison to the rest of the CZ, and the effects of the magnetic
field were modeled as two point masses
lying on a line perpendicular to the axis of the flux tube at the equator. By
further assuming that the star is not tidally locked with the primary, this
nonaxisymmetric contribution to the quadrupole moment exerts an
additional force
onto the companion.
\cite{Applegate89} and \cite{Lanza20} identified two possible scenarios: the
libration model, where the orientation of the flux tube oscillates
around a fixed position, and the circulation model, where the axis of
the flux tube changes in a
monotonic way. These models reduce the energetic requirements by a
factor of $10^2$ to $10^3$ in comparison to the Applegate mechanism,
which is much more restrictive
from an energetic point of view \citep[see e.g.,][]{Brinkworth06, Voelschow16,
Navarrete18, Voelschow18}. Previous models generally require luminosity
variations on the order of 10 per~cent, whereas the improved model of
\citet{Lanza20} reduces the energy requirement by a factor of
$10^2...10^3$.

The transition to predominantly nonaxisymmetric large-scale magnetic fields
in solar-like stars for rapid rotation was investigated by
\cite{Viviani18} with the same model
as that used by \cite{Navarrete20}. They found that the dominant dynamo mode
switches from axi- to nonaxisymmetric at roughly three times the solar rotation
rate. However, this study also showed that the dominant dynamo mode depends
on the resolution of the simulations such that rapidly rotating models at modest
resolutions were again more axisymmetric. In the present study we revisit both
simulations presented in \cite{Navarrete20} with a more detailed analysis
and include an additional
run to explore the importance of (non-) axisymmetric magnetic fields in the
modulation of the gravitational quadrupole moment. The main goal of this
study is to investigate whether dynamo-generated quadrupole moment variations
can lead to the observed period variations and to compare the classical Applegate
machanism with the one of \cite{Lanza20} by means of our simulations.
The dynamo solution is particularly sensitive
to the rotation rate, which is the parameter we focus on in the present
study.

In section \ref{sect:model} we present the model and the methods that we use.
The results are presented in section \ref{sect:results} and a more in-depth
discussion follows in section \ref{sect:discussion}. The conclusions are drawn
in section \ref{sect:conclusions}.


\section{The model}\label{sect:model}

The model employed here is the same as that described in \citet{Kapyla13}
and \citet{Navarrete20}.
We solve the compressible MHD equations in a spherical shell configuration
resembling the solar convection zone with the
{\sc Pencil Code}\footnote{\href{https://github.com/pencil-code}{https://github.com/pencil-code}}
\citep{2021JOSS....6.2807P}.
The equations are
        \begin{align}
              & \frac{\partial \bm{A}}{\partial t} = \bm{u} \times \bm{B} -
                \mu_0\eta\bm{J}, \\
              & \frac{{\rm D}\ln \rho}{{\rm D} t} = - \bm{\nabla}\bm\cdot\bm{u}, \\
              & \frac{{\rm D}\bm{u}}{{\rm D} t} = \bm{g} - 2\bm{\Omega}_0
                \times \bm{u} + \frac{1}{\rho} \left(\bm{J}\times\bm{B} -
                \bm{\nabla}p + \bm{\nabla}\bm\cdot2\nu\rho\bm{\mathsf{S}}\right),\\
              & T\frac{{\rm D}s}{{\rm D}t} = \frac{1}{\rho} \left[-\bm{\nabla}
                 \bm\cdot\left(\bm{F}^{\rm rad} + \bm{F}^{\rm SGS}\right) +
                 \mu_0\eta\bm{J}^2\right] + 2\nu\bm{\mathsf{S}}^2,
        \end{align}
where $\bm{A}$ is the magnetic vector potential, $\bm{u}$ is the velocity field,
$\bm{B} = \bm{\nabla}\times\bm{A}$ is the magnetic field, and
$\bm{J} = \mu_0^{-1}\bm{\nabla}\times\bm{B}$ is the electric current density where
$\mu_0$ is the vacuum permeability. Also, ${\rm D}/{\rm D}t = \partial/\partial t +
\bm{u}\cdot\bm{\nabla}$ is the convective derivative, and $\rho$ is the density.
$\bm{F}^{\rm rad} = - K\bm{\nabla} T$ is the radiative flux and
$\bm{F}^{\rm SGS} = - \chi_{\rm SGS}\rho T \bm{\nabla}s$ is the subgrid-scale
(SGS) flux. The former accounts for the flux coming from the radiative core
to the CZ whereas the latter represents the unresolved turbulent
transport of heat. $K$ is the radiative heat conductivity and $\chi_{\rm SGS}$ is
the turbulent entropy diffusivity. $s$ is the specific entropy, $p$ is the
pressure, and $T$ the temperature. We assume an ideal gas law, that is,
\begin{equation}
        p = (\gamma - 1)\rho e,
\end{equation}
where $\gamma = c_p/c_V = 5/3$ is the ratio of specific heats at
constant pressure and volume, and $e = c_VT$ is the specific internal
energy. The traceless rate-of-strain tensor,
$\bm{\mathsf{S}}$, is defined as
\begin{equation}
        \mathsf{S}_{ij} = \frac{1}{2}(u_{i;j} + u_{j;i}) -
        \frac{1}{3}\delta_{ij}\bm{\nabla}\bm\cdot\bm{u},
\end{equation}
where semicolons denote covariant differentiation. $\bm{g} \propto \hat{\bm r}/r^{2}$
is the
gravitational acceleration. The rotation vector is given by
$\bm{\Omega}_0 = (\cos\theta,-\sin\theta,0)\Omega_0$.

\subsection{Initial and boundary conditions}

The thermodynamic initial state is isentropic. The density
profile follows from hydrostatic equilibrium. The simulations are characterized
by a number of input parameters. These are the energy flux at the bottom, the
angular velocity,
viscosity, magnetic diffusivity, and the radiative and turbulent heat
conductivities and the radial profiles of the latter two. We keep all of them
fixed except for the angular velocity (see Sect. \ref{sec:characteristics}).
Velocity and magnetic fields are initialized with small-scale
low-amplitude Gaussian noise perturbations. These have amplitudes of
$0.25$ m~s$^{-1}$ and $4$~G, respectively.

The computational domain is given by
$0.7R \leq r \leq R$, $\theta_0 \leq \theta \leq
\pi - \theta_0$, $0 \leq \phi \leq 2\pi$, for the radial, latitudinal, and
longitudinal coordinates, respectively, with $\theta_0 = \pi/12$.
Both radial boundaries are impenetrable and stress-free for the flow.
The bottom boundary is a
perfect conductor and the magnetic field at the surface is radial. The upper
boundary follows a blackbody condition. The latitudinal boundaries are
stress-free and perfect conductors. The derivatives of the density and
entropy are zero on both latitudinal boundaries. This implies
that there is no heat flux through these surfaces.

The modeled  star is assumed to have one solar mass
with a convective envelope covering 30\% of the stellar radius.
The simulations labeled as Run~A and Run~B are run3x and run20x
presented in \citet{Navarrete20}. The main difference is that a significantly
longer (over 160 instead of 85 years) time series is available for Run~B.
Furthermore, we perform a more in-depth analysis of both simulations
and include a third simulation, labeled as
Run~C, to ascertain the significance of the
nonaxissymmetric magnetic fields for the gravitational quadrupole moment
(see Sect. \ref{sect:results}). Runs A and B have a resolution of
$128\times256\times512$ in the radial, latitudinal, and azimuthal directions
respectively, and Run~C has a resolution of $128\times288\times512$.

\subsection{Spherical harmonic decomposition}\label{subsect:sph}

To investigate the proposed connection between the nonaxisymmetric component
of the magnetic field and the fluid density, we perform the same decomposition
as in \citet{Viviani18} for the radial magnetic field and density field at
various radial depths (see Figures \ref{fig:Brm1}, \ref{fig:Brm2},
\ref{fig:rhom1}, and \ref{fig:rhom2} for snapshots of the first and second
nonaxisymmetric modes near the surface of the three runs.)
A function $f = f(\theta,\phi)$ can be written as
\begin{equation}\label{eq:sph_harm}
f(\theta,\phi) = \sum_{l=0}^{l_{\rm max}} \sum_{m=-l}^{l} \tilde{f}_{\,l}
^{\,m}(\theta,\phi) Y_l^m(\theta,\phi),
\end{equation}
where
\begin{equation}
\tilde{f}_{\,l}^{\,m} = \int_0^{2\pi}\int_{\theta_0}^{\pi -\theta_0}
f(\theta,\phi)\,Y_l^{m*}\sin{\theta}\,d\theta\,d\phi.
\end{equation}
For the radial magnetic field $B_r(\theta,\phi)$, we impose the
condition
\citep[see][]{Krause80}
\begin{equation}
B_{r,l}^{-m} = (-1)^{m} B_{r,l}^{m*},
\end{equation}
and because the same property applies to the
spherical harmonics $Y_l^m$, we have
\begin{equation}
B_r(\theta,\phi) = \sum_{l=1}^{l_{\rm max}} B_{l,r}^0 Y_l^0 + 2\,{\rm Re}
\left( \sum_{l=1}^{l_{\rm max}}
\sum_{m=1}^l B_{l,r}^m Y_l^r \right).
\end{equation}
The term containing $l = 0$ has been dropped
because it violates solenoidality of the magnetic field.

\subsection{Simulation parameters}\label{sec:characteristics}

Each run is characterized by the Taylor, Coriolis,
fluid and magnetic Reynolds numbers, and fluid, 
SGS and magnetic Prandtl numbers. These are defined as
\begin{gather}
{\rm Ta} = \!\left[ \frac{2\Omega_0(0.3R)^2}{\nu}\right]^2,\
{\rm Co} = \frac{2\Omega_0}{u_{\rm rms} k_1}, \\
{\rm Co}^{(\omega)} = \frac{2\Omega_0}{\omega_{\rm rms}},\
{\rm Re} = \frac{u_{\rm rms}}{\nu k_1},\
{\rm Re_M} = \frac{u_{\rm rms}}{\eta k_1}, \\
{\rm Pr} = \frac{\nu}{\chi_m},\
{\rm Pr_M} = \frac{\nu}{\eta},\
{\rm Pr_{\rm SGS}} = \frac{\nu}{\chi_{\rm SGS}^{\rm m}},
\end{gather}
where $\nu$ is the viscosity, $u_{\rm rms}$ the root-mean-square velocity,
${\rm Co}^{(\omega)}$ is an alternative definition of the Coriolis number based
on the rms vorticity, $k_1 = 2\pi/0.3R$ an estimate of the wavenumber of the
largest convective eddies, $\eta$ the magnetic diffusivity, and
$\chi_{\rm SGS}^{\rm m}$ is the SGS entropy diffusion at
$r=0.85~R_\odot$.


\section{Results}\label{sect:results}

We present the results of three runs, labeled as A, B, and C, with
rotation periods of 8.3, 1.2 days, and 0.8 days, corresponding to 3, 
20, and 30 times the solar rotation rate. We keep all other
sytem parameters fixed.
\begin{table*}
\centering
\caption{Summary of simulation parameters.}\label{tab:parameters}
\begin{tabular}{l|cccccccccccc}
\hline
Run & $\Omega/\Omega_\odot$ & $P_{\rm rot}$ (days) & Ta & Co & Co$^{(\omega)}$
    & Ma & Re & Re$_{\rm M}$ & Pr & Pr$_{\rm M}$
    & Pr$_{\rm SGS}$ & $\Delta$t [yr]\\
\hline
A & 3  & 8.3 & $5.68\times10^7$ & 2.8   & 1.28 & $6.46\times10^{-2}$
  & 66.6 & 66.6 & 58 & 1.0 & 2.5 & 39  \\
B & 20 & 1.2 & $2.52\times10^9$ & 62.4  & 12.4 & $3.29\times10^{-2}$
  & 20.3 & 20.3 & 58 & 1.0 & 2.5 & 138 \\
C & 30 & 0.8 & $5.68\times10^9$ & 139.1 & 22.4 & $2.06\times10^{-2}$
  & 13.6 & 13.6 & 58 & 1.0 & 2.5 & 145 \\
\hline
\end{tabular}
\tablefoot{ $\Omega/\Omega_\odot$ is the rotation rate in
units of the mean solar angular velocity. Co and Co$^{(\omega)}$ are
the Coriolis numbers, Ma is the volume and time averaged Mach number,
Re and Re$_{\rm M}$ are the fluid and  magnetic Reynolds numbers. Pr,
Pr$_M$, and Pr$_{SGS}$ are the fluid, magnetic,
and subgrid scale Prandtl numbers, respectively. $\Delta$t is
the total simulated time. }
\end{table*}
We summarize input and diagnostic quantities that characterize each
simulation in Table \ref{tab:parameters}.

\subsection{Magnetic activity and quadrupole moment evolution}

\begin{figure}
\includegraphics[width=\columnwidth]{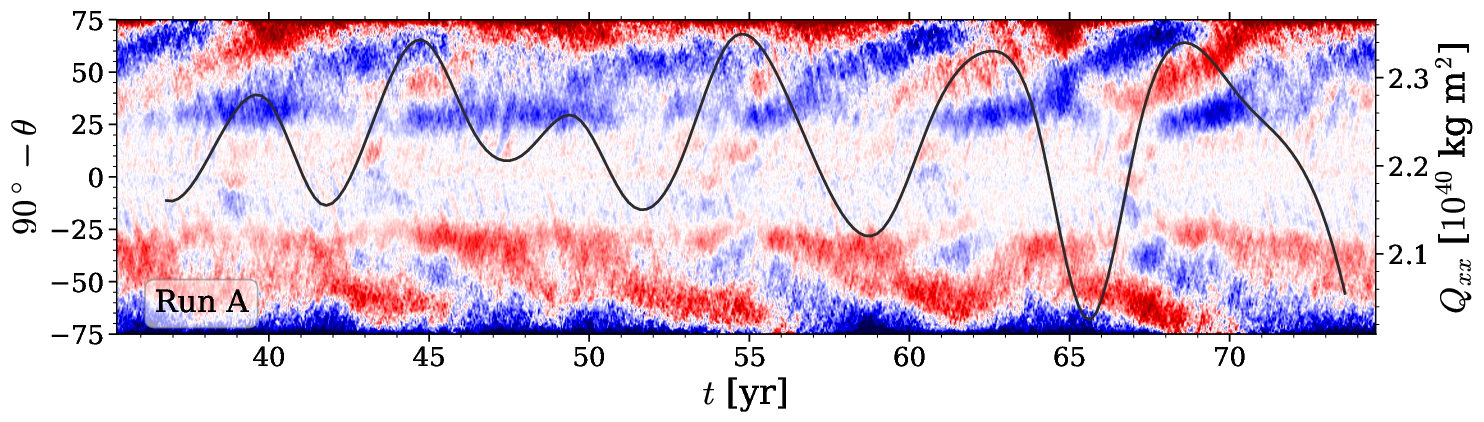}
\includegraphics[width=\columnwidth]{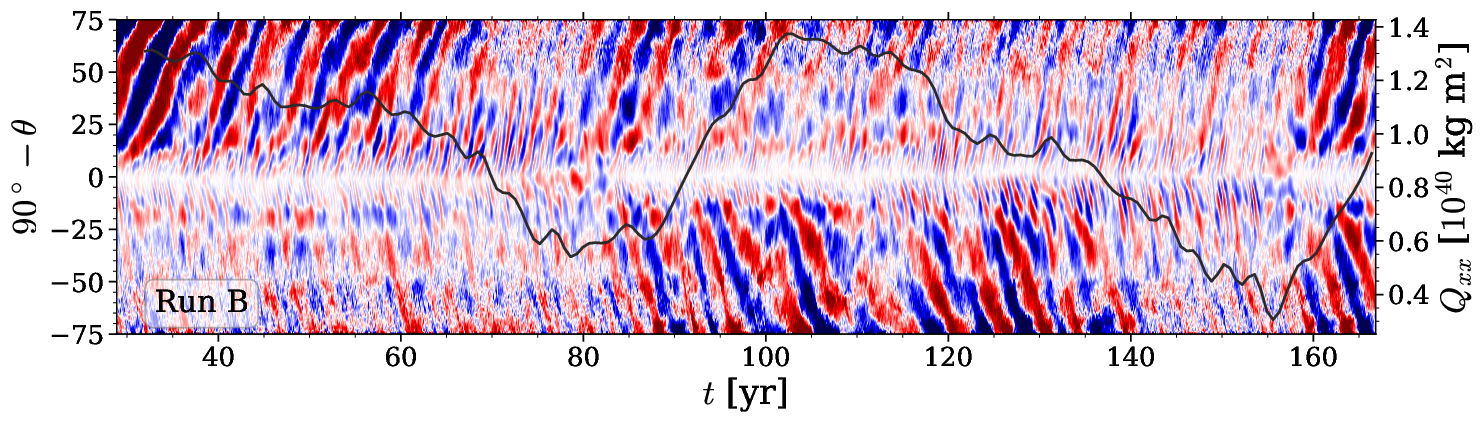}
\includegraphics[width=\columnwidth]{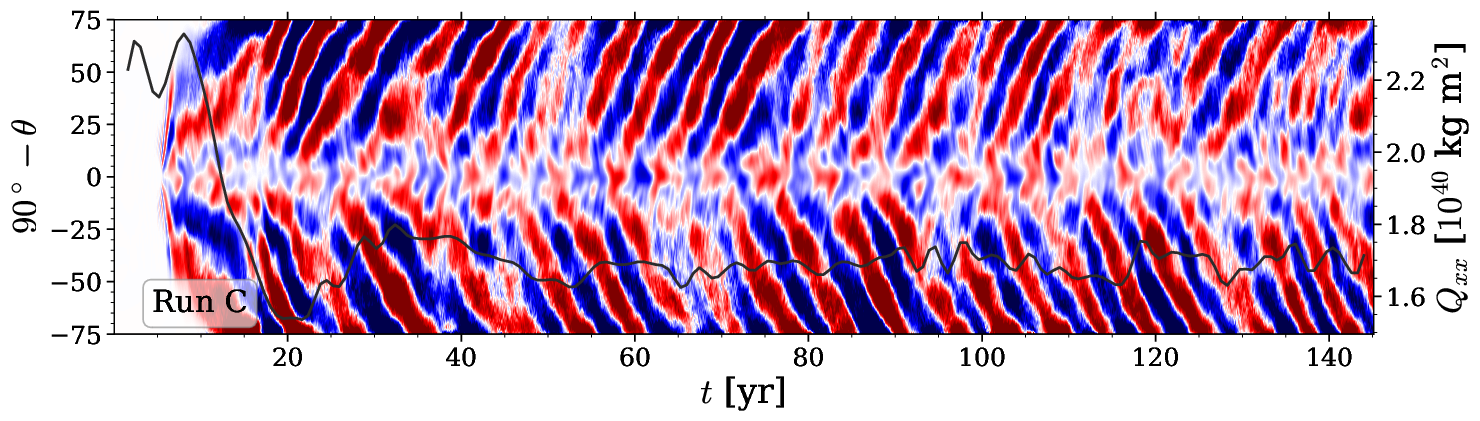}
\caption{Azimuthally averaged radial magnetic field ($\overline{B}_r$) near the
surface of the domain at $r=0.98R$ as a function of latitude and time for 
Run~A (top, $P_{\rm rot}=8.3$~d), Run~B (middle, $P_{\rm rot}=1.2$~d), and Run~C 
(bottom, $P_{\rm rot}=0.8$~d). $Q_{xx}$ for each run is shown from the time
periods where this diagnostic is available. The color scale of  $\overline{B}_r$
in each panel has been clipped at $\pm 8$ kG.}
\label{fig:BrQxx}
\end{figure}

\begin{figure*}
\includegraphics[width=0.33\linewidth]{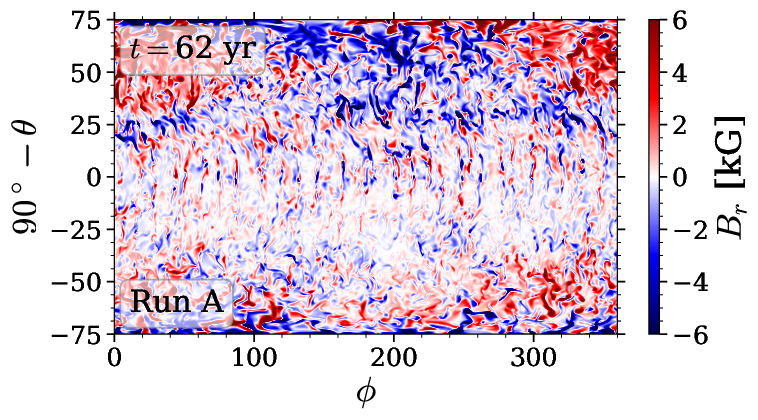}
\includegraphics[width=0.33\linewidth]{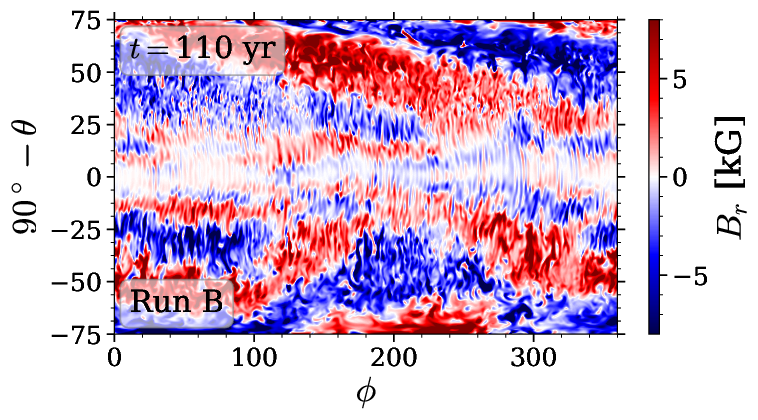}
\includegraphics[width=0.33\linewidth]{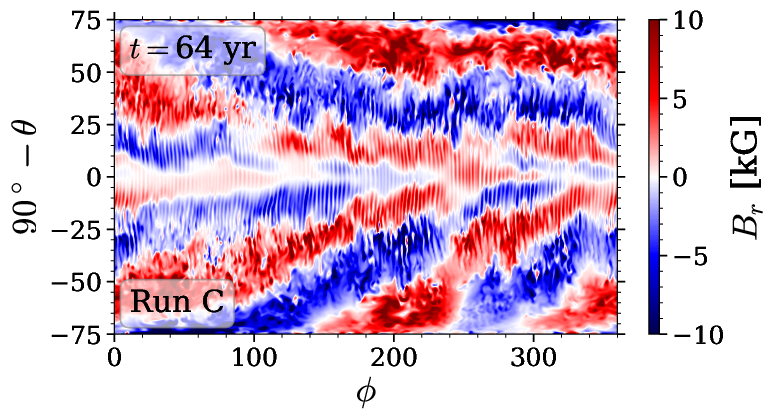}
\includegraphics[width=0.33\linewidth]{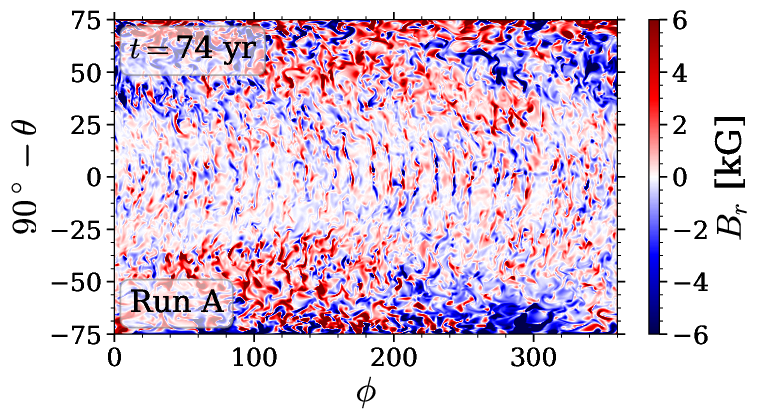}
\includegraphics[width=0.33\linewidth]{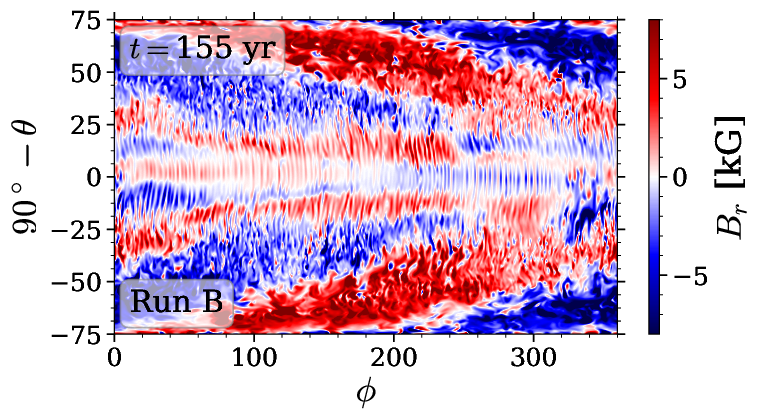}
\includegraphics[width=0.33\linewidth]{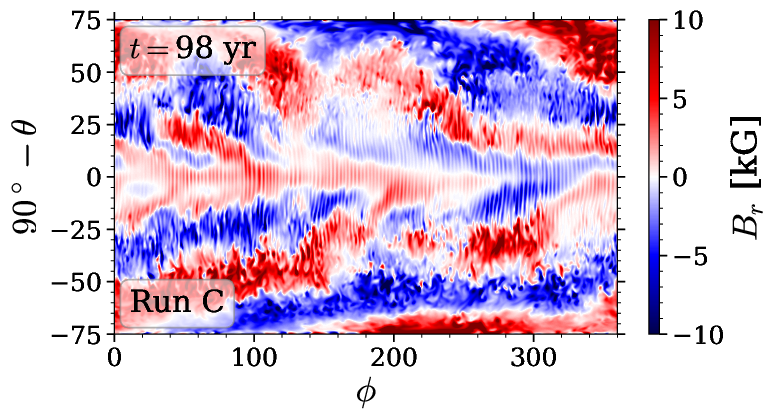}
\caption{Instantaneous radial magnetic field at $r=0.98R$ for each run at two
times. Top (bottom) row corresponds to a maxima (minima) of $Q_{xx}$.}
\label{fig:Br_0.98R}
\end{figure*}

We begin our analysis by comparing the time-dependent diagnostics of magnetic fields
and the gravitational quadrupole moment. in Figure \ref{fig:BrQxx} we show the
azimuthally averaged radial magnetic field ($\overline{B}_r$) near the surface
of the stars at $r/R=0.98$ (color contours), along with the evolution of the
$xx$ component of the gravitational quadrupole moment $Q$ (black lines)\footnote{We note
that the values of $Q_{xx}$ for Runs~A and B differ from \citet{Navarrete20}. This is
because in that study, $I_{zz}$ was erroneously calculated
\citep[see][]{2021MNRAS.504.1676N} . However,
this difference does not change their conclusions.}. The
latter is defined as
\begin{equation}
Q_{ij} = I_{ij} - \frac{1}{3}\delta_{ij}\,{\rm Tr}I,
\label{equ:Qij}
\end{equation}
where
\begin{equation}
I_{ij} = \int\rho({\bm x})x_ix_j{\rm d}V
\label{equ:Iij}
\end{equation}
is the $ij$ component of the inertia tensor expressed in Cartesian coordinates
and $\rho({\bm x})$ is the density. All of the simulations show
low-amplitude variations of $Q_{xx}$ on a timescale of roughly $0.2$
years. These are attributed to sound waves and have a purely
hydrodynamic origin \citep{Navarrete20}. Hence such signals are left
out of the data in this study by working with low-cadence snapshots and
interpolating the data with a cubic interpolator. In Run~A, the variations of
$Q_{xx}$ are small, on the order of $10^{39}...2\times10^{39}$~kg~m$^2$ and 
by visual inspection we estimate that they
have a similar period (roughly 6-8 years) as the
axisymmetric part of the magnetic field. Run B shows a larger amplitude
long-term variation of $Q_{xx}$ that repeats at least once in the data. 
Roughly half of the data for Run B, up to
about 80 years, was presented in \cite{Navarrete20}. The steady
decrease of $Q_{xx}$ between $t\approx30$ to $t\approx85$ years was
interpreted as a
transient due to insufficient thermodynamic and magnetic saturation. However,
with the longer time series we see that the quadrupole moment 
is modulated on a timescale of about 80 years. It also appears
that $Q_{xx}$ is roughly correlated to $\overline{B}_r$: low 
values of the quadrupole moment approximately coincide with times when
$\overline{B}_r$ is weak on both hemispheres ($t=70\ldots85$ and 
$t=140\ldots160$ years, respectively).
The cycle is perhaps starting again at $t=160$ as the
magnetic activity appears to be resuming with a corresponding change in 
$Q_{xx}$. The largest variation of $Q_{xx}$ occurs between
$t=100...155$~years,
with an amplitude of $10^{40}$~kg~m$^2$. It corresponds
to the largest variation of $Q_{xx}$ of the three runs. Overall, the
gravitational quadrupole
moment appears to follow the radial magnetic field strength near the
surface of the star independently of the hemispheric asymmetry.
In contrast to the other two runs, diagnostics for $Q_{xx}$ in Run C
are available starting at $t=0$ yr. The quadrupole moment in this run remains
more or less constant and only starts to decrease after the magnetic field
approaches the saturated regime at $t\approx10$~yr. The
magnetic field back-reacts and re-adjusts the thermodynamic quantities, 
such as density, after which $Q_{xx}$ settles to a state with smaller
variations around a mean value
of $1.70\times10^{40}$~kg~m$^2$. These variations are about half of
those in Run~A.
As seen in Fig.~\ref{fig:BrQxx}, the axisymmetric part of $B_r$ is similar in
Runs~A and C, but clearly different in Run~B. In the first two, $\overline{B}_r$
migrates toward higher latitudes in a regular fashion and in the latter
the dynamo is more hemispheric such that activity alternates between 
both hemispheres seemingly
every 50~to~60~years. 
Overall, cycles of $Q_{xx}$ in Runs A and C follow more closely the
polarity reversals of the magnetic field. There are also such cycles in
Run~B. However, they are hidden by the longer cycle that follows the migration
of the hemispheric component of the magnetic field.

\begin{table}
\caption{Summary of some quantities of interest corresponding to data
  shown in Fig.~\ref{fig:BrQxx}.
}
\label{tab:energies}
\begin{tabular}{l | c c c c c c}
\hline
Run & $Q_{xx}^{\rm{max}} $ & $ Q_{xx}^{{\rm min}}$ &
$ \overline{B}_{r}^{\rm max} $ & $ \overline{B}_{r}^{\rm min}$ &
$E_{\rm mag,total}^{\rm max}$ & $E_{\rm mag,total}^{\rm min}$ \\
\hline
A & 2.25 & 1.94 & 8.65 & -9.70  & 2.98 & 1.26 \\ 
B & 1.38 & 0.315 & 24.7 & -19.7 & 3.98 & 2.28 \\
C & 1.80 & 1.54 & 16.8 & -13.8  & 2.32 & 1.30 \\
\hline
\end{tabular}
\tablefoot{$Q_{xx}$ is in units of 10$^{40}$ kg m$^{2}$ and $\overline{B}_r$ in
units of kG. $E_{\rm mag,total}$ is the volume-averaged magnetic energy in units
of $10^{5}$~J~m$^{-3}$.}
\end{table}

In Fig.~\ref{fig:Br_0.98R} we show instantaneous snapshots of the radial magnetic
field near the surface of the three runs at the times of interest,
that is, maxima
(minima) of $Q_{xx}$ at top (bottom) row.
These correspond to $t=62,74$~yr for Run~A, $t=110,155$~yr for Run~B,
$t=64,98$~yr for Run~C.
In Run~A, a predominantly $m=1$ large-scale mode is seen at high latitudes
but this is subdominant
in comparison to the axisymmetric ($m=0$) component (see Appendix
\ref{sec:decomposed}). In Run~B
there is a dominant $m=1$ mode on the northern hemisphere, while a
predominantly $m=2$ mode dominates on the southern hemisphere at the
maximum. At the minimum of $Q_{xx}$
(middle panel in the lower row), $B_r$ is symmetric with respect to the
equator with a dominating $m=1$ mode on both hemispheres. These large-scale
structures cover the entire hemispheres from the equator to the latitudinal
boundaries. This is qualitatively similar to Run~H of \citet{Viviani18}
and Run~C of \citet{Cole14}. As suggested by Fig. \ref{fig:BrQxx},
a similar pattern
repeats for Run~B at $t\approx36$~yr and $t\approx75$~yr but in opposite
hemispheres. Run~C is similar to Run~B in the sense that the large-scale
nonaxisymmetric structures are promiment over a large region, but the
$m=1$ and $m=2$ modes appear to be similar in strength. These modes
appear to alternate between the hemispheres
but the variations in the gravitational quadrupole moment are
weak in this case in comparison to Run B. However, the low-order
nonaxisymmetric
fields in Run C are of the same order of magnitude as the $m=0$ mode whereas
in Run B the $m=1$ mode is clearly stronger than the axisymmetric fields.
A possible explanation of the smaller cycles of $Q_{xx}$ found in Run~C
can be attributed to the dynamo solution. Slow rotators tend to produce dominant
$m=0$ modes. Faster rotators typically show a predominant $m=1$ mode
although sometimes the $m=0$ component can become dominant again at
very high rotation rates if the convection is only weakly
supercritical \citep{Viviani18}. It is plausible that this is
happening in our Run~C. These results suggest that a dominant
nonaxisymmetric magnetic field with
hemispheric asymmetry is associated with the strongest variations of $Q$.

\begin{figure*}
\includegraphics[width=0.33\linewidth]{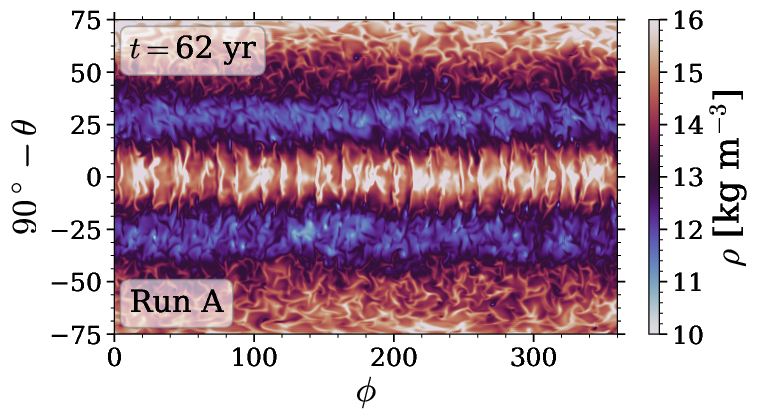}
\includegraphics[width=0.33\linewidth]{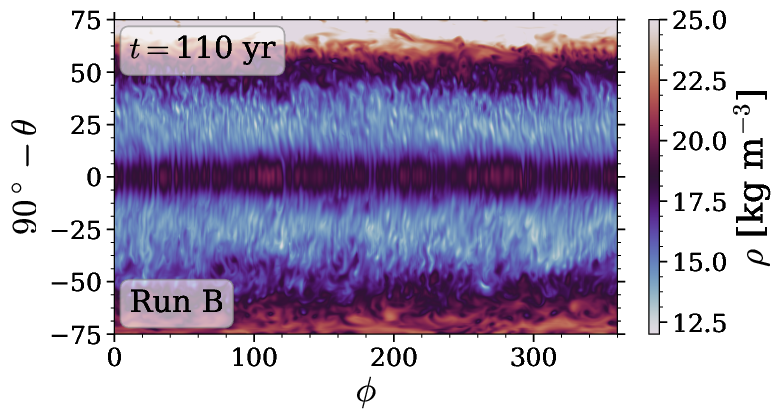}
\includegraphics[width=0.33\linewidth]{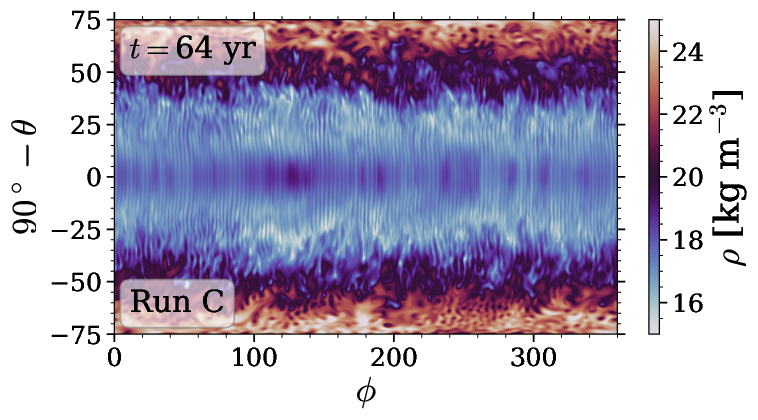}
\includegraphics[width=0.33\linewidth]{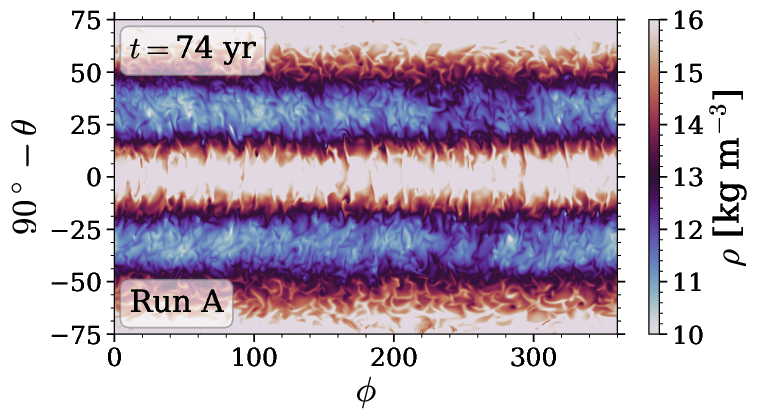}
\includegraphics[width=0.33\linewidth]{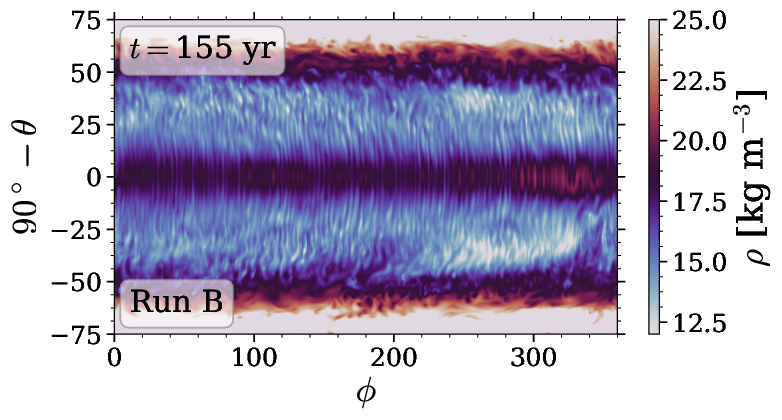}
\includegraphics[width=0.33\linewidth]{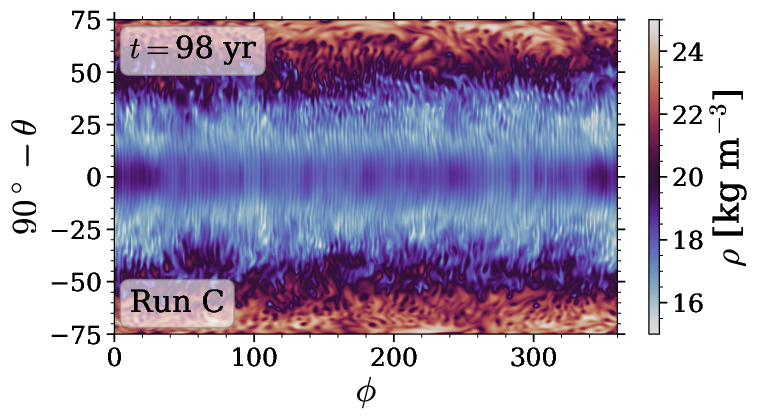}
\caption{Instantaneous snapshots of density at $r=0.98R$ for each run.}
\label{fig:rho}
\end{figure*}

\subsection{Density variations and structural changes due to magnetic fields}

The variations of the gravitational quadrupole moment are related to changes
in the mass distribution within the star as can be seen from
Eqs.(\ref{equ:Qij}) and (\ref{equ:Iij}).
Snapshots of the density from all three runs near the surface of the star are
shown in Fig. \ref{fig:rho}. As
before, the shown times correspond to maxima (top row) and minima (bottom
row) of $Q_{xx}$. Snapshots of the $m=1$ and $m=2$ modes of density are show
in the Appendix~\ref{sec:decomposed}.

In Run~A there is an overall change in density between the two
times. At $t=62$ yr (top panel), when the gravitational quadrupole
moment is larger, there are no noticeable large-scale nonaxisymmetric
features,
whereas when $Q_{xx}$ is at a minimum ($t=74$ yr), weak nonaxisymmetric
features appear.
This can be seen from the two blue stripes around $\theta = \pm 30^\circ$ where
the overall density decreases with patches of increased (decreased) density
around $\phi=270^\circ$ ($\phi=60^\circ$). Closer to the poles and
near the equator the average density increases but no clear
nonaxisymmetric features
are present. In Run~B we identify a few characteristics. First, when the
quadrupole moment is larger at $t=110$~yr, there is a clear asymmetry with
respect to the equator, such that the density is larger close to the north pole. As
$Q_{xx}$ decreases, the asymmetry disappears, and
nonaxisymmetric structures become visible at
$230^\circ<\phi<340^\circ$ and $\theta=\pm40^\circ$. As the magnetic field
changes its configuration from one that is dominated by an $m=1$ mode only
at the northern hemisphere to predominantly $m=1$ on both hemispheres
(see Fig. \ref{fig:Br_0.98R}), the density field reacts and also changes to a
nonaxisymmetric configuration with a corresponding change in the gravitational
quadrupole moment. In contrast, large-scale density variations in Run~C between
the two
times are clearly weaker. The density field at the surface remains symmetric with
respect to the equator as well as in the azimuthal direction.

\begin{figure*}
\includegraphics[width=0.33\linewidth]{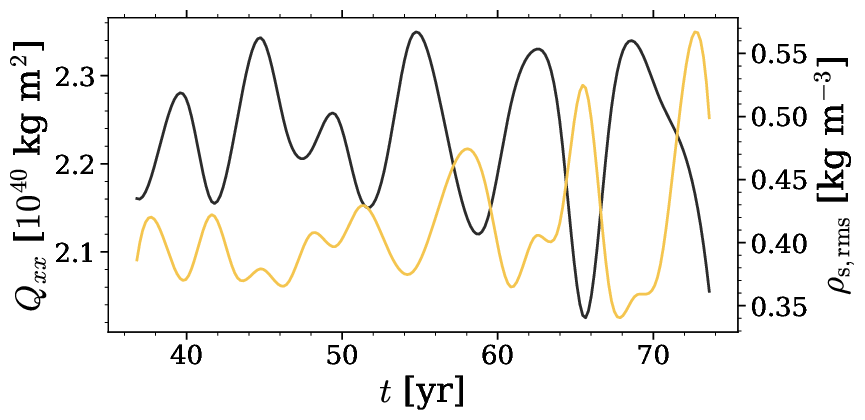}
\includegraphics[width=0.33\linewidth]{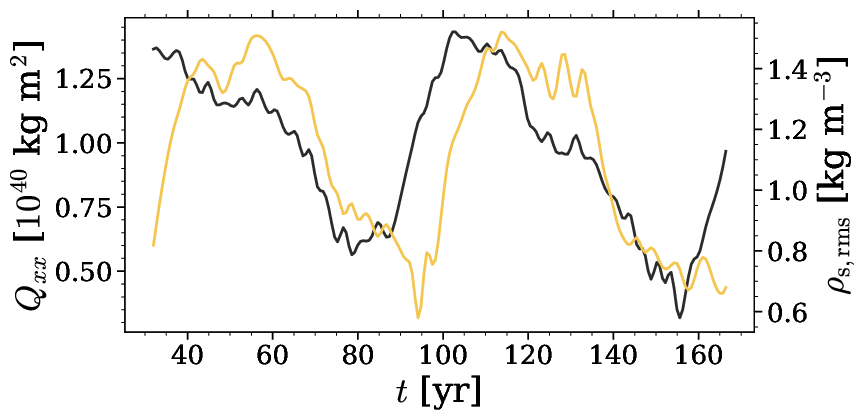}
\includegraphics[width=0.33\linewidth]{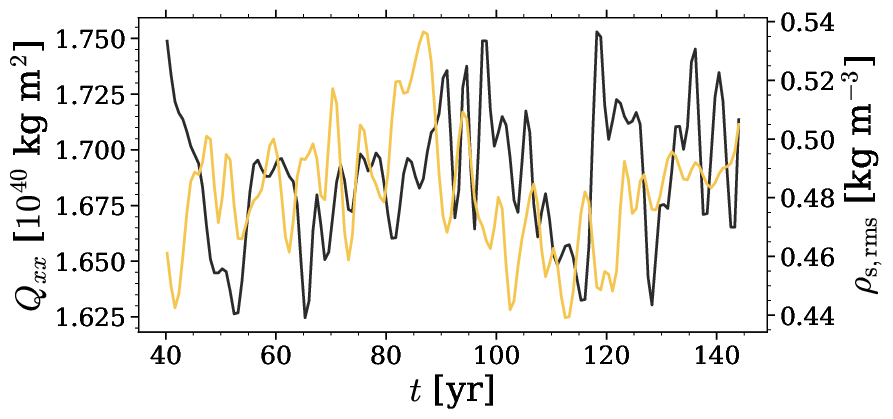}
\includegraphics[width=0.33\linewidth]{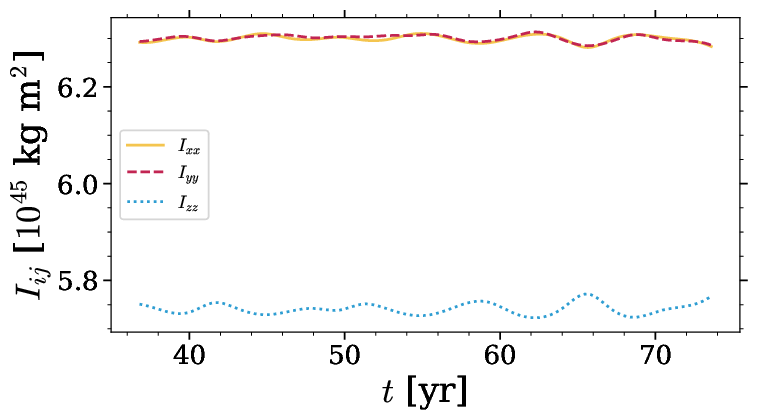}
\includegraphics[width=0.33\linewidth]{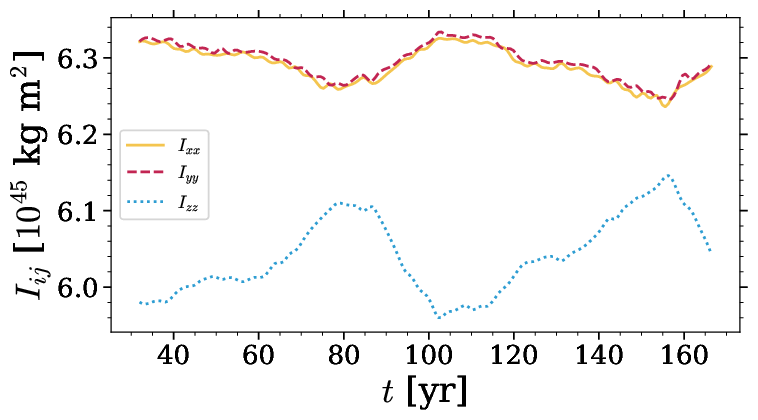}
\includegraphics[width=0.33\linewidth]{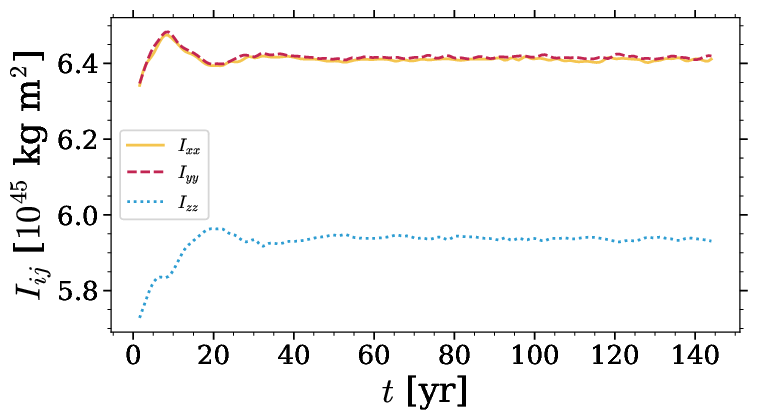}
\caption{Top row: $Q_{xx}$ (black) and rms value of the equatorically
  asymmetric part of density $\rho_{\rm s,rms}$ (yellow)
  according to Eq.~(\ref{equ:rhoasym2}) as functions of time.
  Bottom row: time evolution of the diagonal components of the inertia tensor.
  First, second, and third columns correspond respectively to runs A, B, and C.}
\label{fig:rhosrms}
\end{figure*}

To investigate the importance of equatorial asymmetry on the
gravitational quadrupole moment, we define the equatorially asymmetric
part of density as
\begin{equation}
        \rho_{\rm asym}(r,\theta,\phi) = \frac{\rho(r,\theta,\phi) - \rho(r,-\theta,\phi)}{2}
\label{equ:rhoasym}
\end{equation}
and compute the root mean square according to
\begin{equation}
        \rho_{\rm s,rms} =
        \left(\langle\rho_{\rm asym}^2\rangle_{\theta\phi}\right)^{1/2}.
        \label{equ:rhoasym2}
\end{equation}
The time evolution of $\rho_{\rm s,rms}$, together with $Q_{xx}$, is
shown in the top row of
Fig.~\ref{fig:rhosrms}. In Run~A (left panel) there is an anticorrelation
between the two. However, in the
case of Run~B (middle panel) there is a positive correlation between the two
but with an apparent time delay.
The rms value of $\rho_{\rm s,rms}$ lags behind $Q_{xx}$ by
roughly $10$~yr for example near the extrema between $80$ and $100$~yr.
In Run C both the variations of the density and $Q_{xx}$ are weak. It is
also less clear whether a correlation between the two exists.
The variations of $\rho_{\rm s,rms}$ are between 6 to 10 times larger
in Run~B than in Runs~A and C, indicating that the former is in a different
dynamo regime where the magnetic field is more strongly coupled to the
density field leading to stronger quadrupole moment variations.

The differences in density with respect to the equator should translate in
variations of the moment of inertia aligned with the rotational axis of the star,
namely, $I_{zz}$. The lower row of Figure \ref{fig:rhosrms} shows the evolution
of the three components of the inertia tensor that contribute to $Q_{xx}$,
which is computed from
\begin{equation}\label{eq:qxxexp}
        Q_{xx} = I_{xx} - \frac{1}{3}\left(I_{xx}+I_{yy}+I_{zz}\right).
\end{equation}
The vertical component $I_{zz}$ is always smaller than the two other
components. In Runs~A and C
all components of $I_{ij}$ have comparable variations, whereas in Run~B (middle
panel) the variations of $I_{zz}$ are significantly larger. This coincides
with larger variations of $\rho^{m=1,2}$ in this run.
In Run~B a maximum of $Q_{xx}$ coincides with a minimum of $I_{zz}$. This
corresponds to the star rotating slightly faster at maxima (minima) of $Q_{xx}$
($I_{zz}$).

\begin{figure*}
\includegraphics[width=0.33\linewidth]{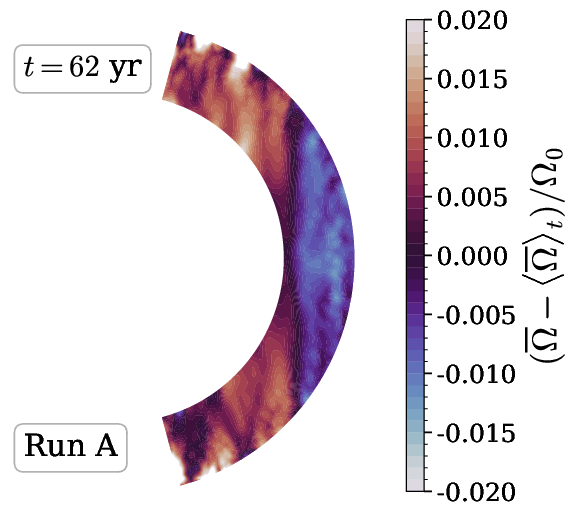}
\includegraphics[width=0.33\linewidth]{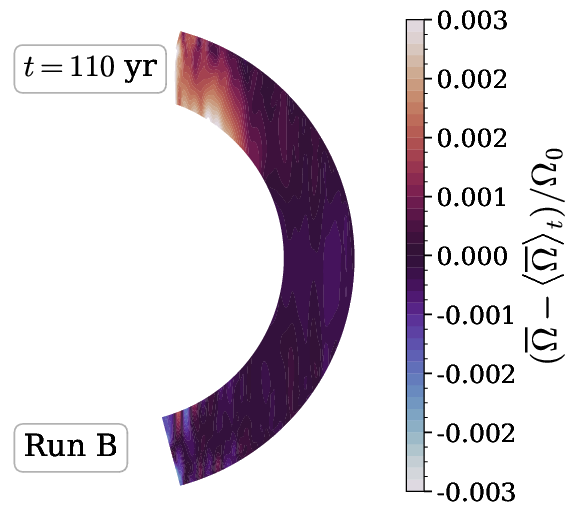}
\includegraphics[width=0.33\linewidth]{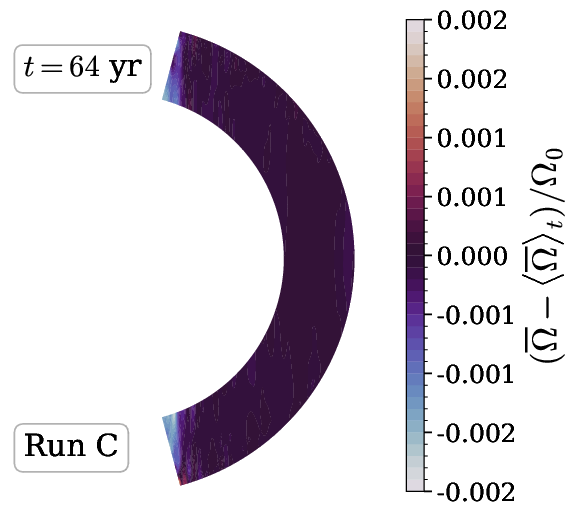}
\includegraphics[width=0.33\linewidth]{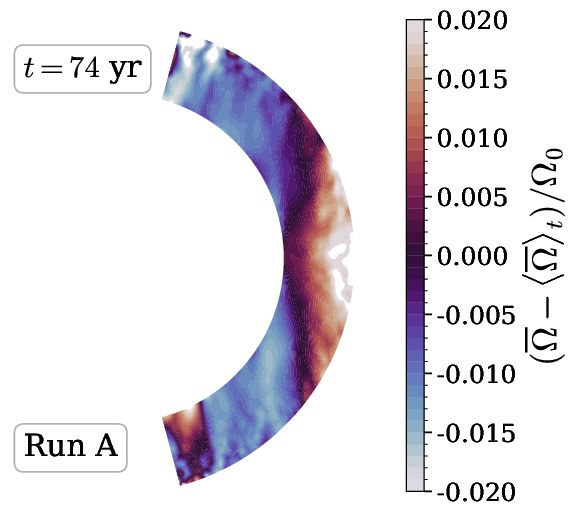}
\includegraphics[width=0.33\linewidth]{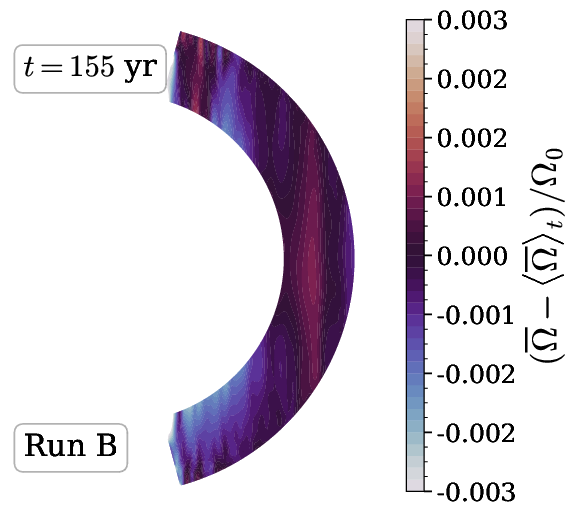}
\includegraphics[width=0.33\linewidth]{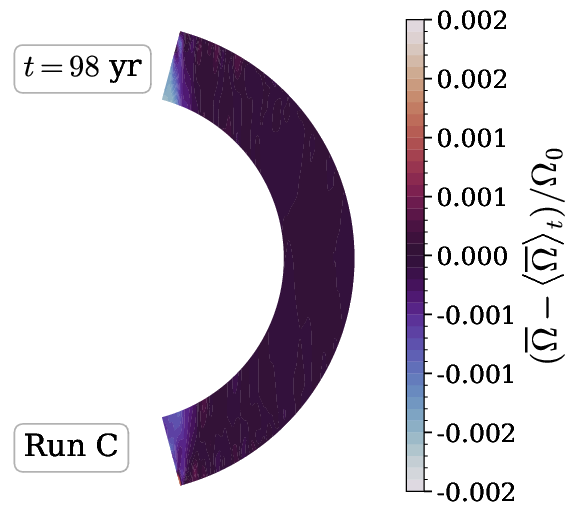}
\caption{Deviations from the time-averaged mean angular velocity
  $(\overline{\Omega}-\langle\overline{\Omega}\rangle_t)$ normalized
  by the angular velocity of the frame of reference $(\Omega_0)$ for
  each run from the times indicated in the legends.}
\label{fig:OmegaProfileAvg}
\end{figure*}

To see such differences, we compute the azimuthally averaged rotation
profiles
\begin{equation}
  \overline{\Omega}(r,\theta) = \Omega_0 + \frac{\overline{u}_\phi(r,\theta)}{r\sin\theta}
\end{equation}
for all runs, average them over time, and show the deviations from such averages
during the two times of interest in Fig.~\ref{fig:OmegaProfileAvg}.
There are minor
differences in the rotation profiles of Runs~A and B between maxima and
minima of $Q_{xx}$, while almost no differences are observed in
Run~C. Runs~A and B have a larger difference between angular
velocities of polar and equatorial
regions at a minima of $Q_{xx}$ (lower panels), but an accelerated northern
pole is seen in the top panel of Run B. A large difference in
differential rotation implies that the star would deform and adopt an
ellipsoidal shape as a consequence of the
centrifugal force, adding a further contribution to the
quadrupole moment. However, as we have fixed boundary conditions, we cannot
model such a reaction.
All of our runs show a solar-like rotation profile with equatorial regions
rotating faster than the poles as a consequence of Coriolis numbers above
the transition region from antisolar to solar-like differential rotation.
This transition occurs around $\rm{Co} \gtrsim 1$ \citep[see
  e.g.,][]{Gastine14, Kapyla14} whereas the rotation profile approaches 
solid body rotation for rapid rotation
\citep[e.g.,][]{Viviani18,Kapyla21}.

\begin{table*}
\centering
\caption{Correlation coefficients between the time series of the quadrupole
  moment and magnetic energy, mean surface density and quadrupole moment,
and mean surface density and magnetic energy}.
\label{tab:corr}
\begin{tabular}{l|ccccccc}
\hline
Run & $Q_{xx}|E_{\rm mag}^{m=0}$ & $Q_{xx}|E_{\rm mag}^{m=1}$
    & $Q_{xx}|E_{\rm mag}^{m=2}$ & $\overline{\rho}_{\rm surf}|Q_{xx}$
    & $\overline{\rho}_{\rm surf}|E_{\rm mag}^{m=0}$ 
    & $\overline{\rho}_{\rm surf}|E_{\rm mag}^{m=1}$ 
    & $\overline{\rho}_{\rm surf}|E_{\rm mag}^{m=2}$ \\
\hline
A   & -0.41 &  0.54 &  0.13 & -0.67 &  0.20 & -0.18 & -0.07 \\
B   &  0.38 & -0.64 &  0.66 & -0.94 & -0.40 & -0.53 & -0.57 \\
C   &  0.26 &  0.22 & -0.34 & -0.40 &  0.25 &  0.11 & -0.11 \\
\hline
\end{tabular}
\end{table*}

\begin{figure*}
\includegraphics[width=0.33\linewidth]{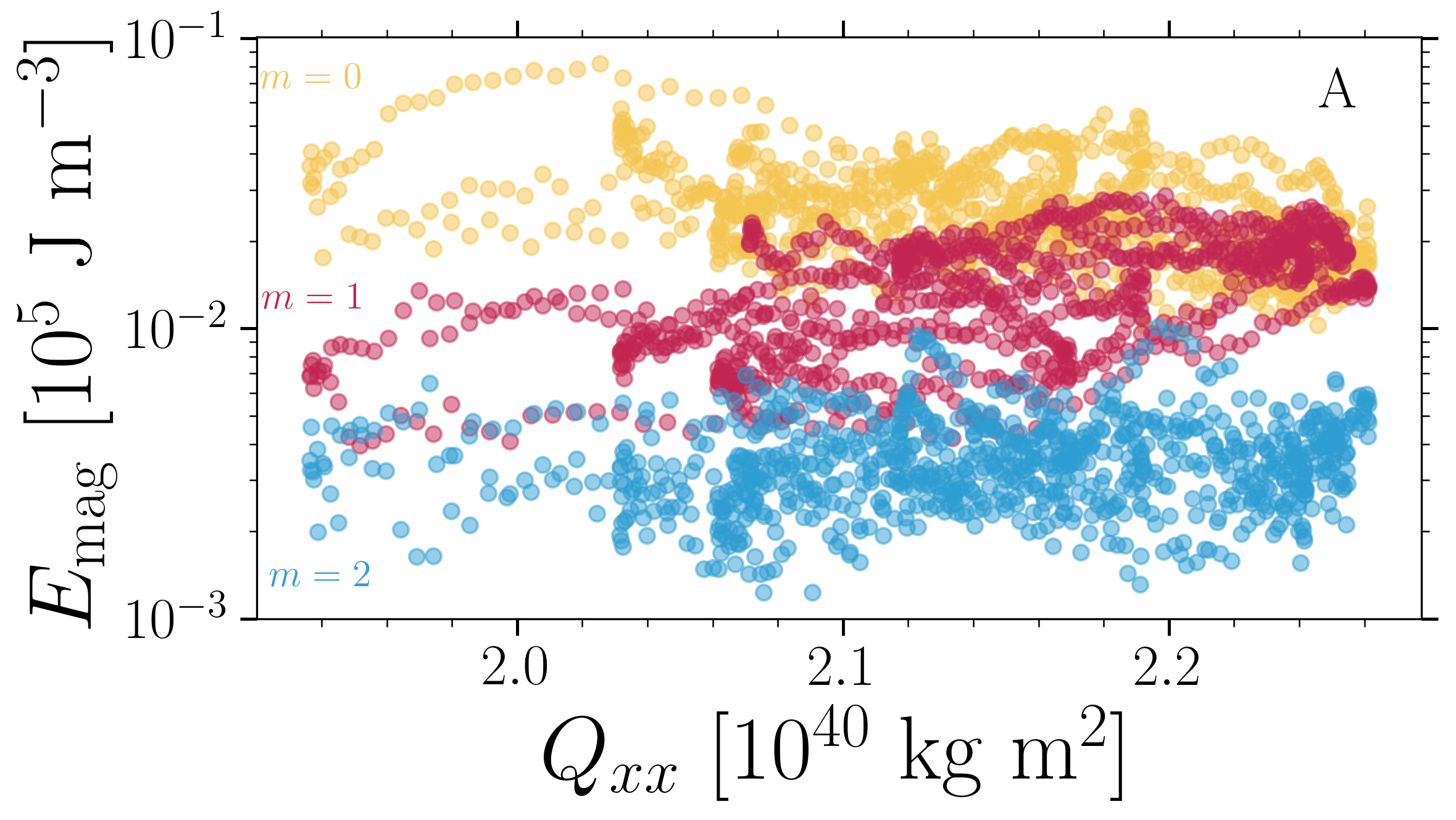}
\includegraphics[width=0.33\linewidth]{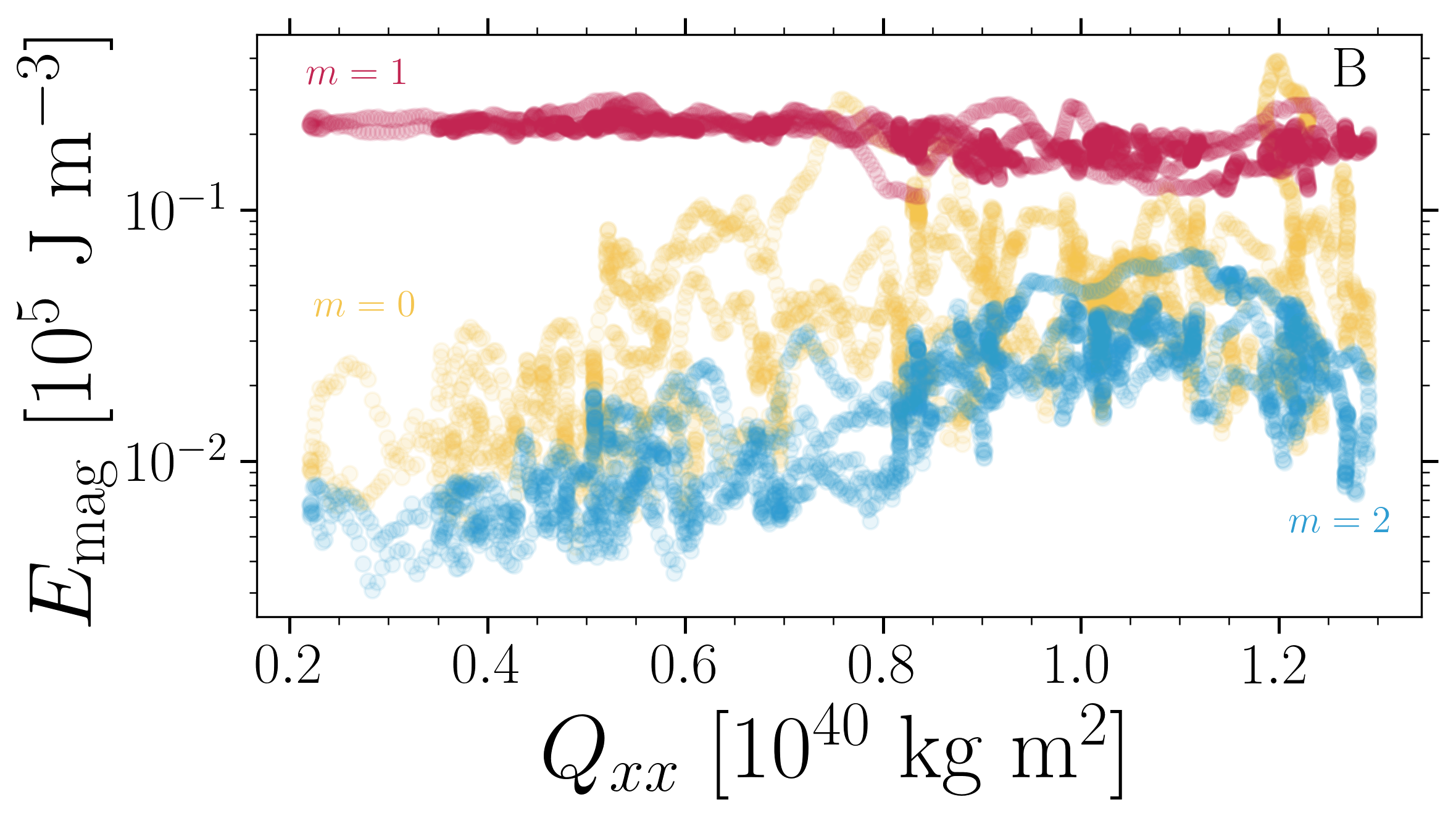}
\includegraphics[width=0.33\linewidth]{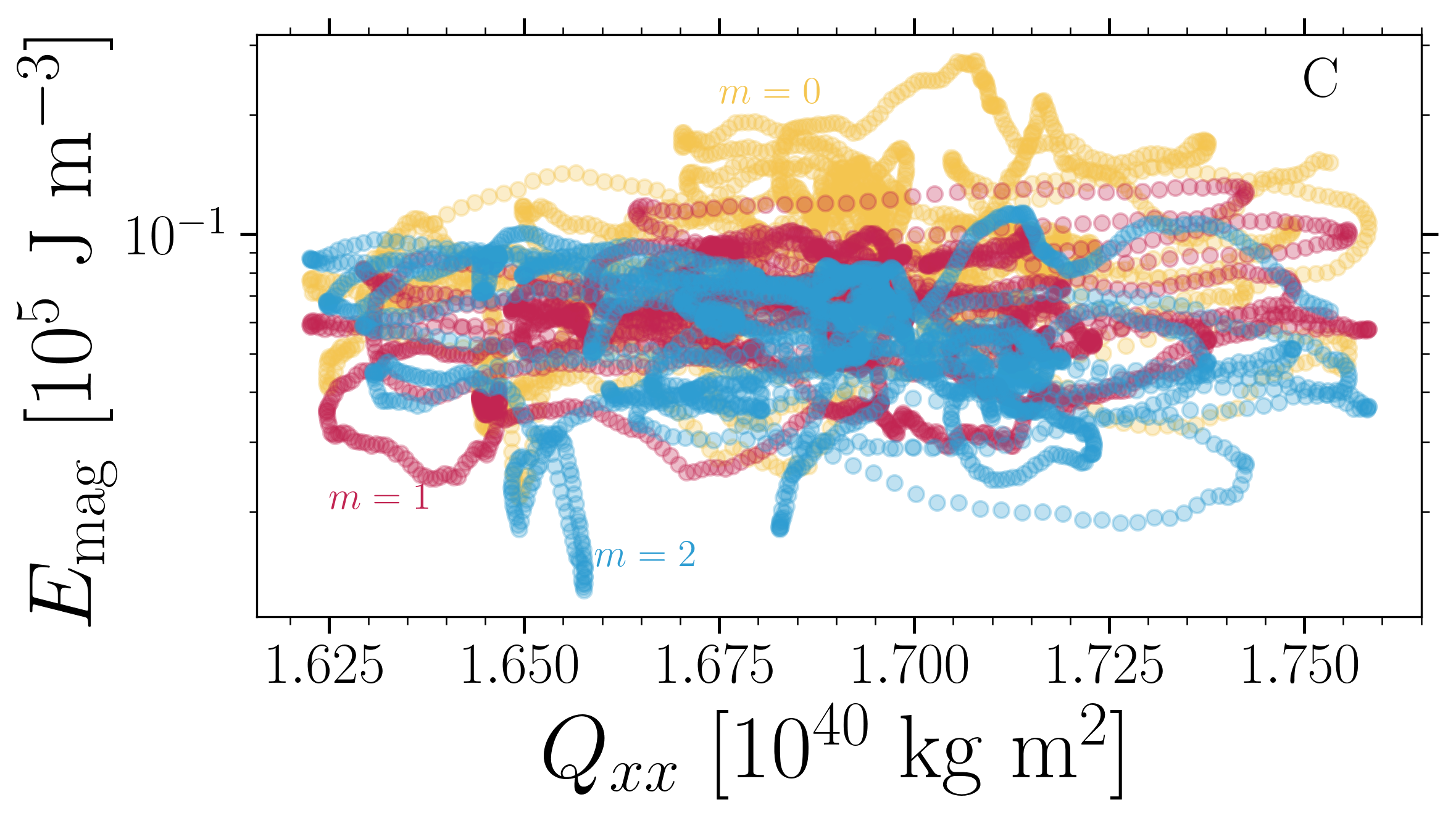}
\includegraphics[width=0.33\linewidth]{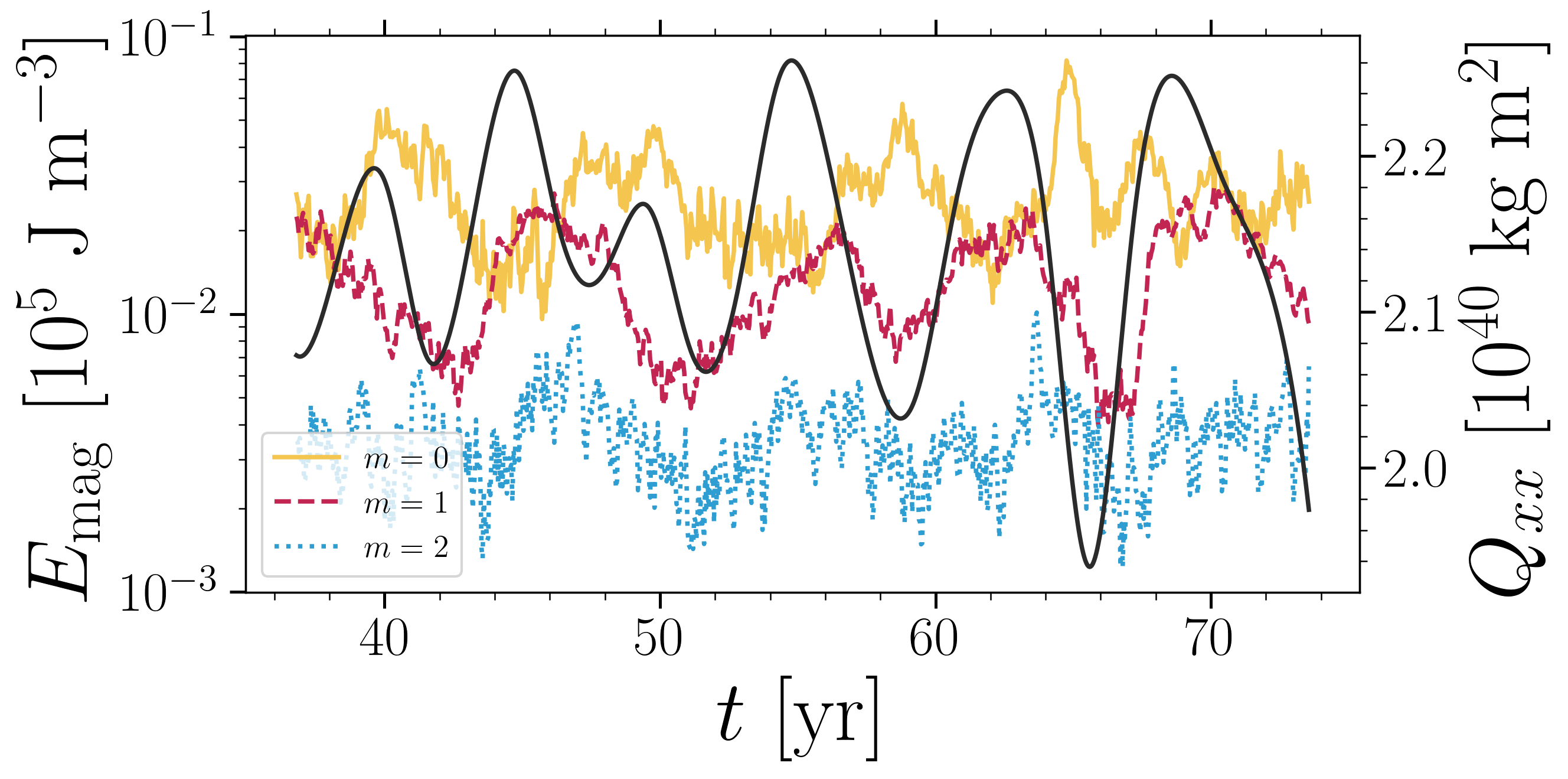}
\includegraphics[width=0.33\linewidth]{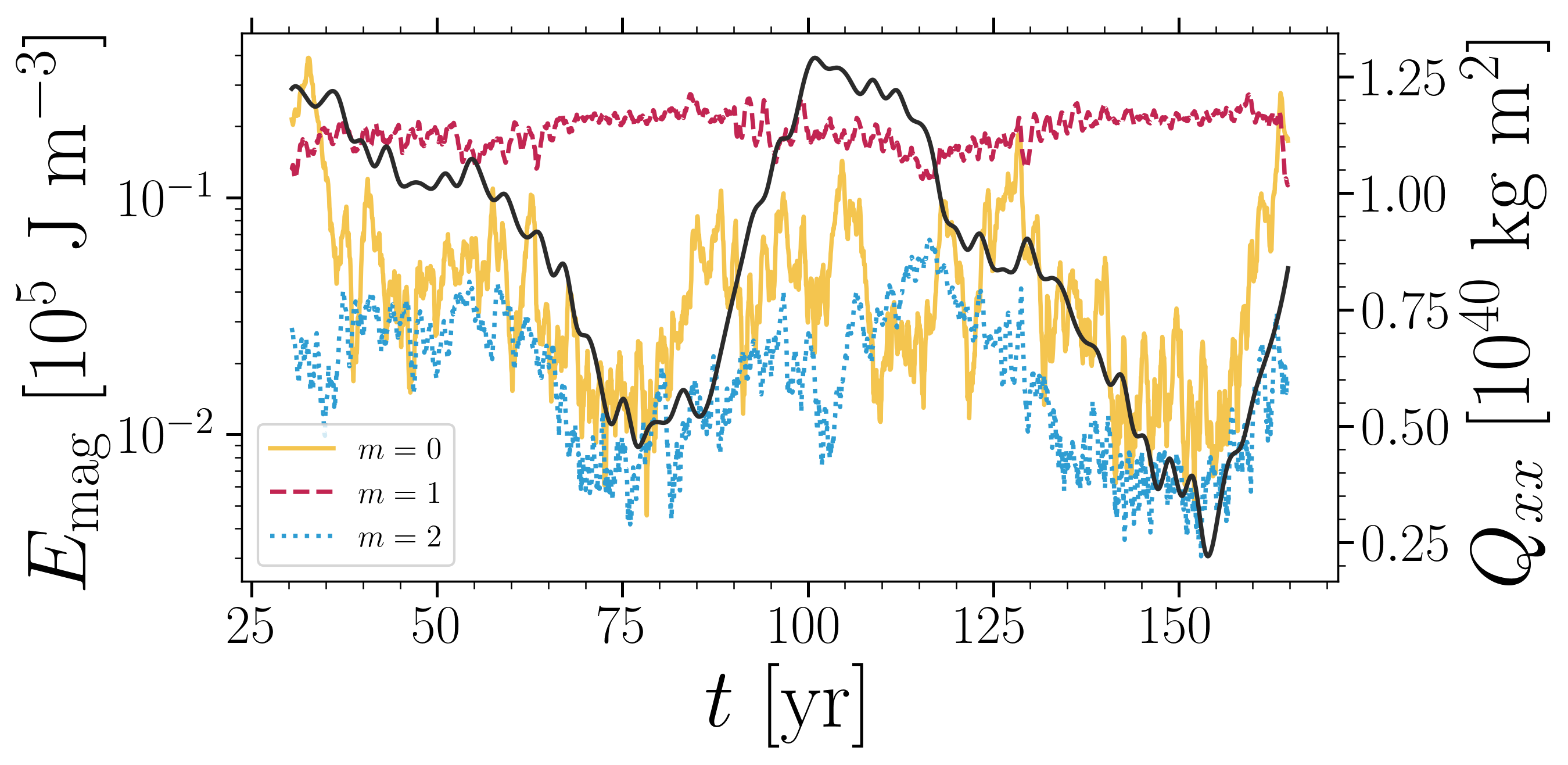}
\includegraphics[width=0.33\linewidth]{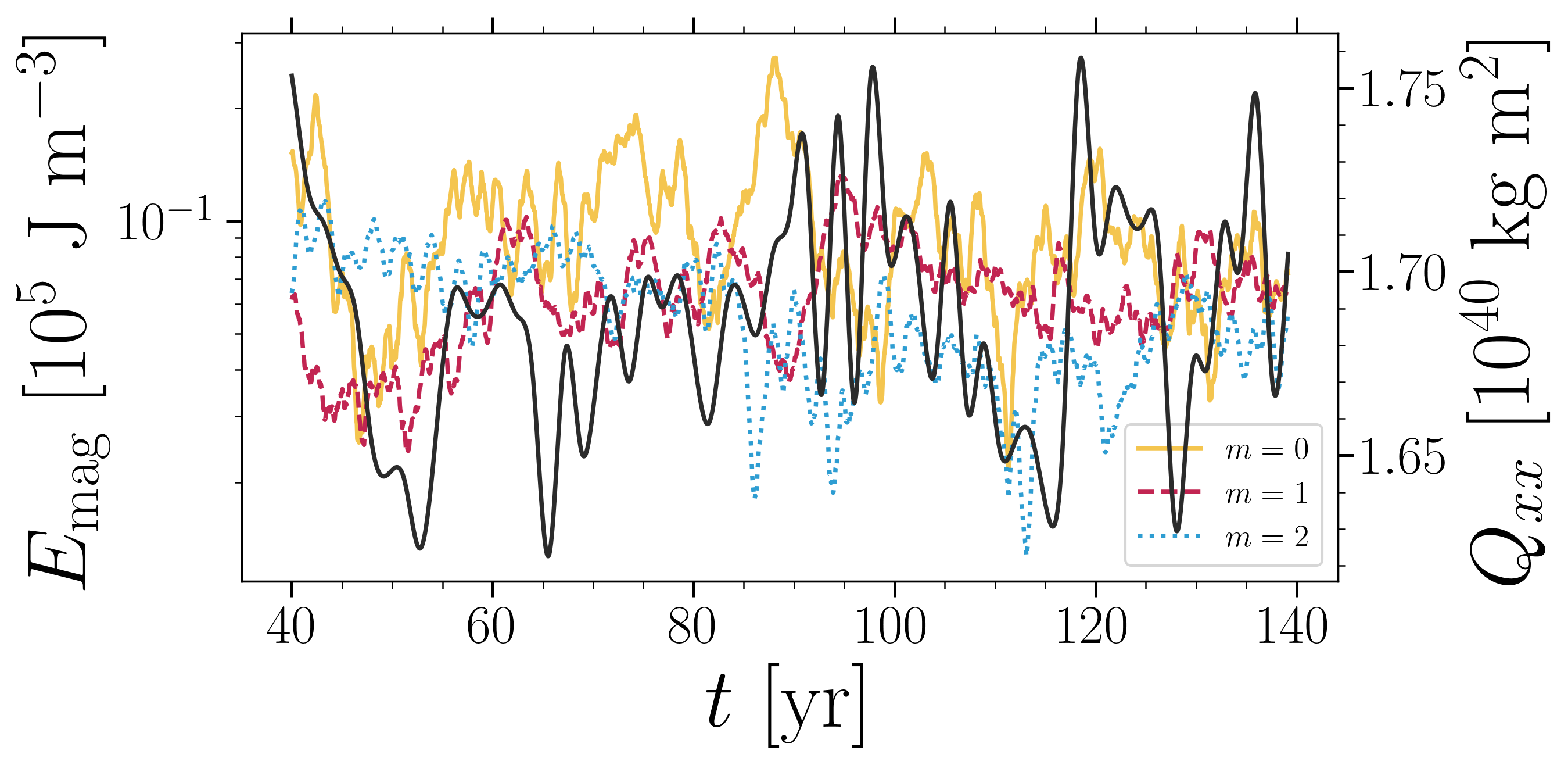}
\caption{Quadrupole moment, density, and moments of inertia. Top row: scatter
    plot of the magnetic field energy and the quadrupole
    moment. Bottom row: Time evolution of the gravitational quadrupole moment
    (black line) together with the magnetic energy contained in the axisymmetric mode ($m=0$,
yellow), as well as the first ($m=1$, red) and second ($m=2$, blue) non
-axisymmetric modes. Run~A, Run~B, and Run~C are shown in the left, middle, and
right columns respectively.}
\label{fig:QxxEmag}
\end{figure*}

The energies of the axisymmetric and first two nonaxisymmetric modes of $B_r$
are shown in the lower row of Fig. \ref{fig:QxxEmag} and a scatter plot between
$B_r^{m=i}$ and $Q_{xx}$ is shown in the upper row. In Run~A (first
column) the axisymmetric
$m=0$ is dominant, which seems to be anticorrelated with $Q_{xx}$. There are
short episodes where the $m=0$ and $m=1$ modes have comparable energies.
The latter is correlated to $Q_{xx}$. In Run~A, the $m=2$ mode
is always subdominant. From the scatter plot we see that no noticeable increase
in magnetic energy is needed to reach a larger quadrupole moment.
The situation in Run~B is different as there is a
persistent $m=1$ mode that dominates throughout the simulation with minor
fluctuations in its energy. The axisymmetric mode seems to be correlated to
$Q_{xx}$. This is because, as explained in the previous section, it produces
equatorially asymmetric density fluctuations that modulate the moment of inertia
aligned with the rotation axis of the star. The second nonaxisymmetric mode
($m=2$) is as strong as the axisymmetric mode and also correlates with $Q_{xx}$.
Larger quadrupole moment values are related to higher energies in the $m=0$ and
$m=2$ modes. In Run~C all modes have similar energy levels, with $m=0$ being
slightly stronger than the other two. The scatter plot reveals that there is no
relation between the magnetic energy and the quadrupole moment.

To quantify the relation between magnetic field modes and quadrupole moment,
we perform a correlation analysis. We use the Pearson correlation coefficient
to study the linear correlation between the density and magnetic fields.
The coefficient between a paired data $(x,y)$ of $n$ pairs is defined as
\begin{equation}\label{eq:correlation}
        x|y = \frac{\sum_{i=1}^n(x_i-\bar{x})(y_i-\bar{y})}
                      {\sqrt{\sum_{i=1}^n(x_i-\bar{x})^2}
                       \sqrt{\sum_{i=1}^n(y_i-\bar{y})^2}},
\end{equation}
where $\bar{x}$ and $\bar{y}$ are the sample mean and $-1 \leq x|y \leq 1$.
A value of $x|y = 1$ ($x|y = -1$) implies perfect (anti-)correlation.

The correlations between the gravitational quadrupole moment and magnetic energy
are calculated using the data presented in Fig. \ref{fig:QxxEmag} and in this
case the barred quantities in Eq. \ref{eq:correlation} represent time averages
of $Q_{xx}$ and $E_{\rm mag}$, whereas $x_i$ and $y_i$ are the time-dependent
quantities of $Q_{xx}$ (calculated over the whole volume) and $E_{\rm mag}$
(calculated over the surface layer). This is shown in the second, third,
and fourth columns
of Table \ref{tab:corr}. In general, we find correlations higher than
0.5 for the $m=1$ mode in Run A and for the $m=1$ and $m=2$
modes in B.

Next, we compute the correlation coefficients between the mean surface 
density $\overline{\rho}_{\rm surf}$ and quadrupole moment, and 
$\overline{\rho}_{\rm surf}$ and the magnetic energies. These are shown in
the last four columns of Table~\ref{tab:corr}. In each run we find that
the values of $\overline{\rho}_{\rm surf}|Q_{xx}$ are relatively large.
This is due to the direct relation between the density distribution
and $Q_{xx}$. The outer regions of the stars are particularly
important due to the $x^2$ dependence of the inertia tensor.
As noted earlier, the dynamo solution in Run~B alternates between the
two hemispheres and so does the density field. As $\overline{\rho}_{\rm surf}$
increases, so does $I_{zz}$ and thus $Q_{xx}$ decreases (see
Eq.~\ref{eq:qxxexp}). Overall, we see the clearest correlations
in Run~B.

The magnetic energy increases with the rotation rate as expected, but this does
not suffice to explain the variations in the gravitational quadrupole moment.
Overall, Run~B has the largest magnetic energy and Run~A and C have
comparable energies (see Table \ref{tab:energies}). However, Run~C has
smaller fluctuations in $Q_{xx}$ which
can be attributed to the fact that the variations of the
magnetic field do not result in significant density perturbations relevant
for the quadrupole moment.

\subsection{Azimuthal dynamo waves}

\begin{figure*}
\includegraphics[width=0.33\linewidth]{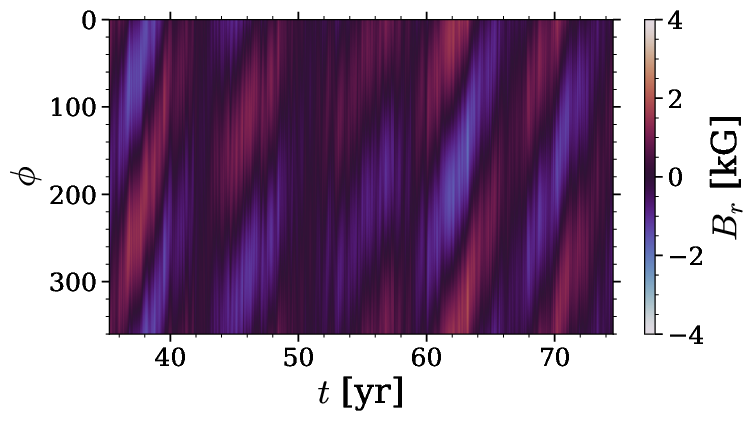}
\includegraphics[width=0.33\linewidth]{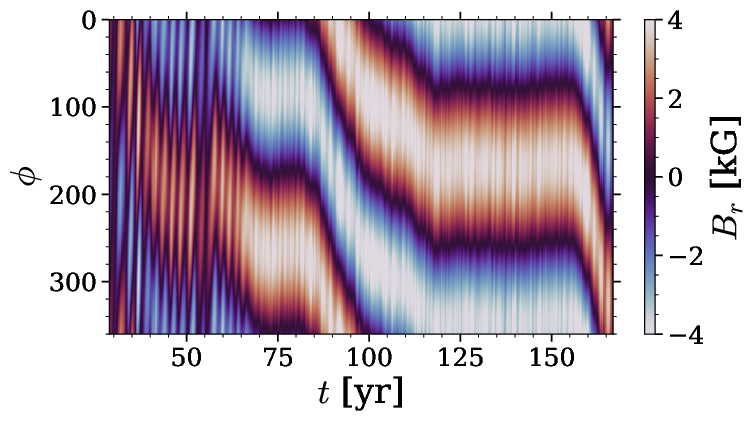}
\includegraphics[width=0.33\linewidth]{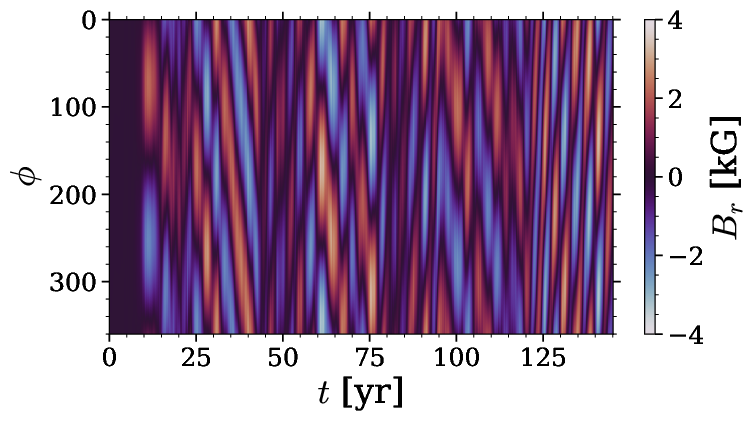}
\includegraphics[width=0.33\linewidth]{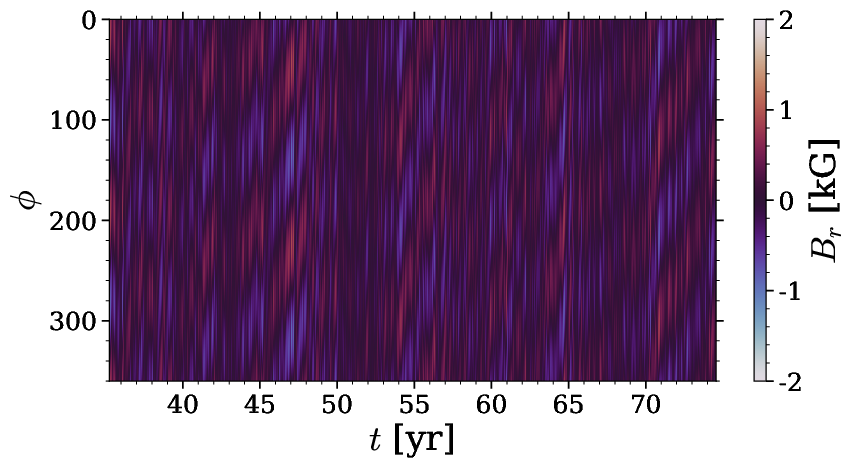}
\includegraphics[width=0.33\linewidth]{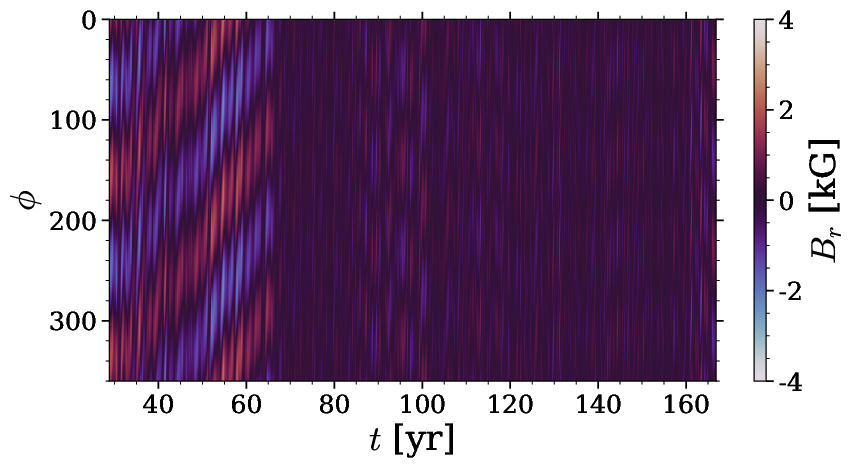}
\includegraphics[width=0.33\linewidth]{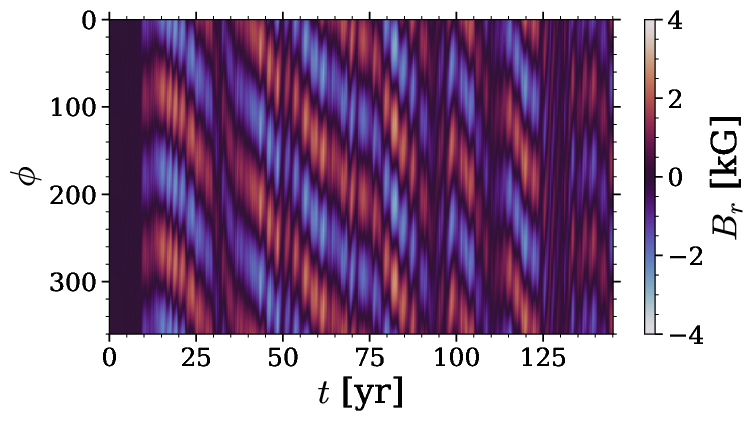}
\caption{Migration of the $m=1$ (top row) and $m=2$ (bottom row) modes of the
radial magnetic field at the surface along the longitudinal direction. for Run~A
(left), Run~B (middle), and Run~C (right).}
\label{fig:ADW}
\end{figure*}

\begin{figure*}
\includegraphics[width=0.33\linewidth]{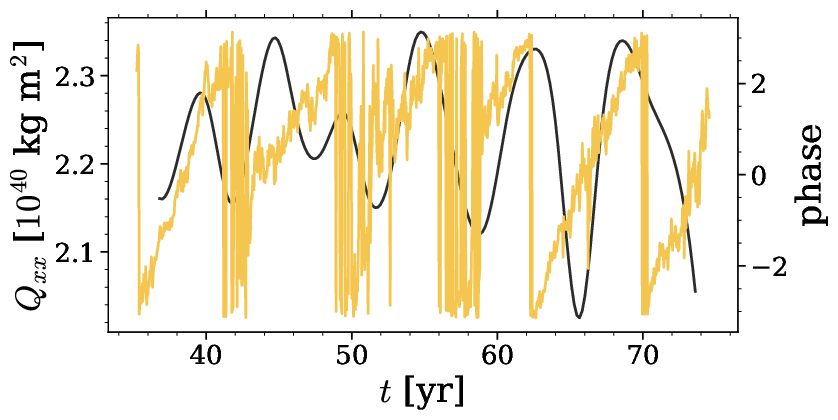}
\includegraphics[width=0.33\linewidth]{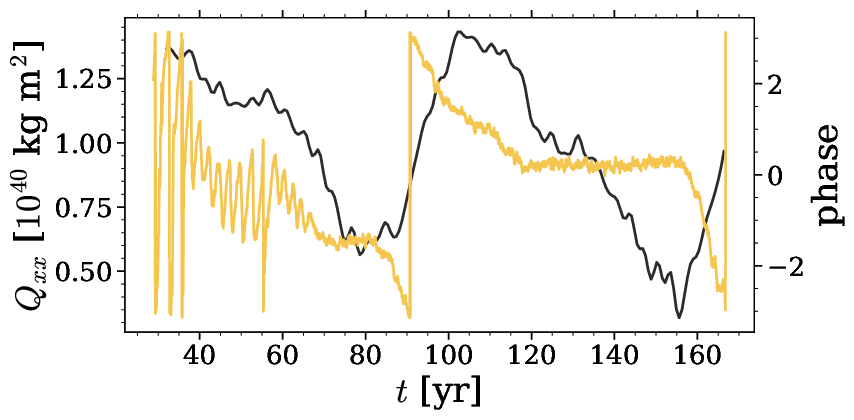}
\includegraphics[width=0.33\linewidth]{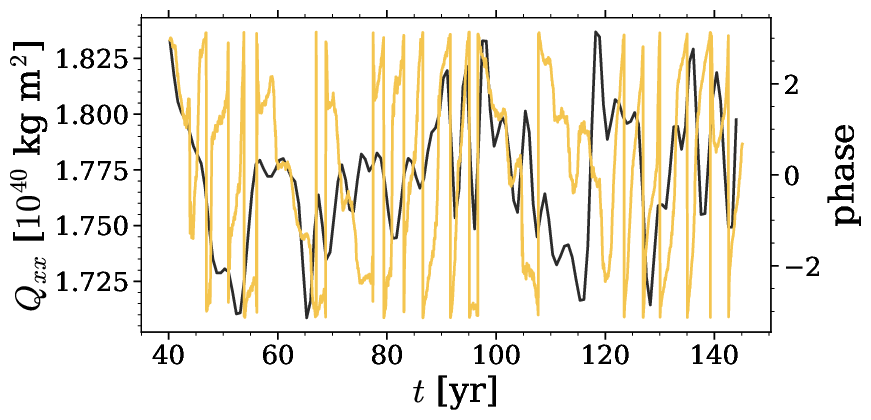}
\includegraphics[width=0.33\linewidth]{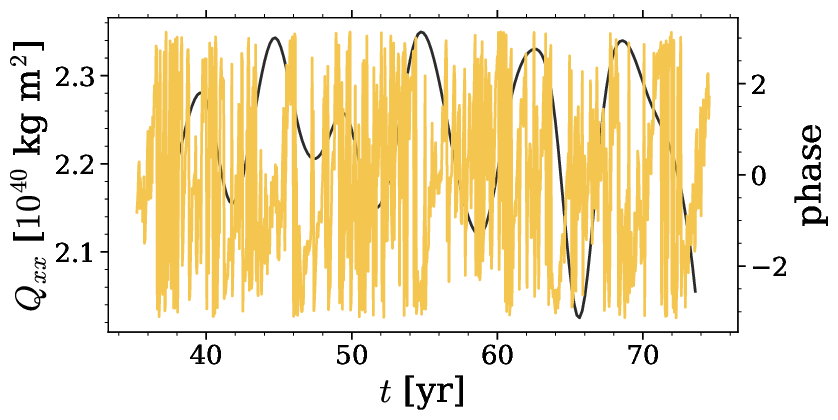}
\includegraphics[width=0.33\linewidth]{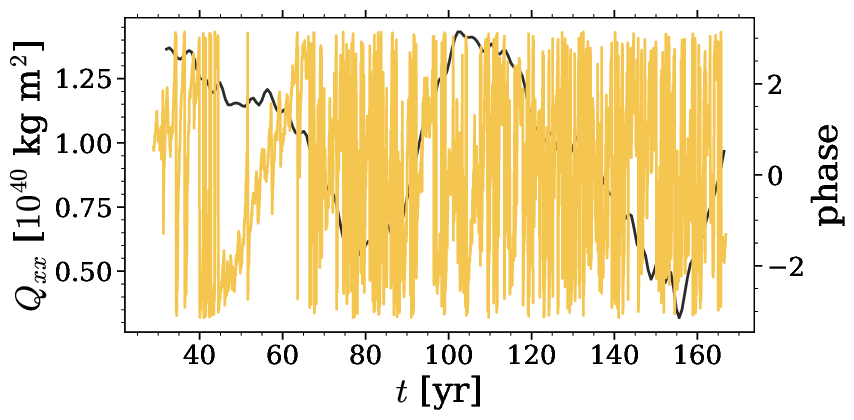}
\includegraphics[width=0.33\linewidth]{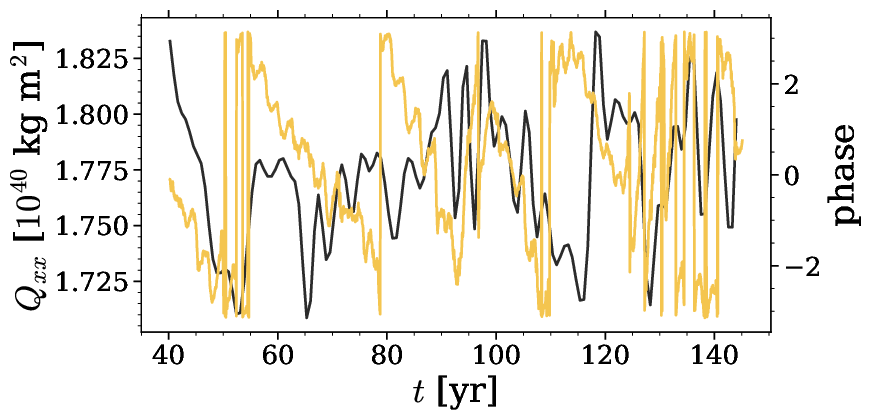}
\caption{Time evolution of the phase of the $m=1$ (top row) and $m=2$ modes
(bottom row) of radial magnetic field at $\phi=180^\circ$ for Run~A
(left panels), Run~B (middle panels), and Run~C
(right panels) in yellow. The black line corresponds to $Q_{xx}$.}
\label{fig:phase}
\end{figure*}

Azimuthal dynamo waves (ADWs) are magnetic structures that migrate in the azimuthal
direction. ADWs can be prograde or retrograde and their propagation is
unaffected by the differential rotation of the star \citep[see
  e.g.][]{Krause80, Cole14, Viviani18}.
The periods of ADWs are usually on the order of a few years for slow
rotators and of tens
of years for faster rotators \citep[][]{Viviani18}. This is comparable
to the period of the quadrupole
moment variations of the present study. We therefore briefly study ADWs.
We take $m=1$ and $m=2$ modes of the radial magnetic field near the surface of
each run at latitude $+60^\circ$ and show the evolution of the field in
Fig.~\ref{fig:ADW} and its phase in Fig.~\ref{fig:phase}.

Run~A has an $m=1$ ADW that migrates in the retrograde direction with
a period of $\sim8$ years in particular after around $t=60$~yr. Meanwhile, the
$m=2$ ADW has
a similar amplitude with no clear periodicity. In Run~B there is a clear $m=1$
wave which migrates in the prograde direction. However, between 115 and 145~yr
the migratory process is stalled for nearly thirty years.
The ADW in Run~B has a period of roughly $\sim80...100$~years. There is an $m=2$
ADW but it is only
noticeable during the first $\sim40$ years of the simulation.
In Run~C there are two equally strong ADWs. Both migrate in a prograde way with
the difference that the period of the $m=1$ wave is shorter than the one of
$m=2$.

The evolution of the phase of the $m=1$ mode correlates well with the
evolution of the gravitational quadrupole moment in Runs~A and B,
whereas the second
nonaxisymmetric mode is noisier and does not correlate with $Q_{xx}$. The phase
of $B_r^{m=1}$ does not show any particular evolution in Run~C, while for $m=2$
the phase is constantly decreasing which is uncorrelated with $Q_{xx}$.
The cases of Runs~A and B point to an underlying relation between a
nonaxisymmetric dynamo mode and the gravitational quadrupole moment
evolution.

\section{Discussion and implications}\label{sect:discussion}

The Applegate mechanism \citep{Applegate92}
is based on the redistribution of angular momentum throughout the star
due to the centrifugal force.
More recently, \citet{Lanza20} presented a new mechanism where the centrifugal
force is no longer needed. In this work the quadrupole moment is constant
in the frame of reference of the magnetically active star due to a
time-invariant nonaxisymmetric magnetic field modeled by a single flux
tube. The companion star experiences a time-varying nonaxisymmetric
gravitational quadrupole moment due to the assumption that the active
star is not tidally locked.

In our simulations we compute the gravitational quadrupole moment in the
rotating frame of reference of the star. The simulations thus
provide a test whether magnetic activity can significantly influence 
stellar structure. We stress that the physical process occurring here
is not the classical Applegate mechanism which is based on the
centrifugal force, and which is not included in our simulations.
The results described in Sect. \ref{sect:results} show that the connection
between magnetic fields and gravitational quadrupole moment is quite complex.
It is due to the asymmetry of the magnetic field with respect
to the equator rather than due to nonaxisymmetry, which is particularly
noticeable
in Run~B.
The three simulations we present differ only in the rotation rate of the star
and yet they present different scenarios of quadrupole moment variations.

\subsection{Behavior of the dynamo itself}\label{subsec:dynamo}

Simulations of stellar magneto-convection have shown that dynamo
solutions depend mainly on the rotation rate of the star.
For example, \citet{Viviani18}
studied the transition from axi- to nonaxisymmetric magnetic fields as a
function of rotation and found that the transition to nonaxisymmetry
occurs for $\Omega \gtrsim 1.8 \Omega_\odot$.
However, \citet{Viviani18} found that at sufficiently rapid rotation, the
magnetic
field returns to a predominantly axisymmetric configuration if the
resolution is not high enough. A similar sequence is also observed in the
simulations described in this paper. Run~A is at a regime where the
axisymmetric mode is slightly stronger than the first nonaxisymmetric mode.
The rotation rate in Run~B is 6.7 times greater than in Run~A and there
the $m=1$ mode dominates, while $m=0$ and $m=2$ are comparable.
In Run~C, with 10 times faster rotation than in Run~A, the dominant
mode is again $m=0$. If the resolution
was to be increased, corresponding to higher Reynolds numbers, it is possible
that a nonaxisymmetric solution with stronger quadrupole moment variations
would be recovered.

Besides resolution effects, it is possible that Run~B is in a parameter regime where
hemispherical dynamos are preferred; see, for example,
\citet{Grote&Busse00}, \citet{Busse02}, and \citet{Kapyla10}. \citet{Brown20}
reported a cyclic single-hemisphere dynamo in a simulation of a fully convective
star, and mentioned that such dynamos are present in other  simulations with
similar parameters. \citet{Kapyla21} presented a set of simulations of
fully convective
stars where in a single run short periods of hemispheric dynamo action were
seen, but even in this case the dynamo is predominantly present on both
hemispheres. It is important to explore this parameter
regime in more detail as it can potentially help to address the question of whether the
ETVs in PCEBs have a magnetic origin. If this is the
main ingredient, however, it would imply that PCEBs that show variations in
the ${\rm O-C}$ diagram are in this particular regime.

\subsection{Classical Applegate mechanism}\label{subsec:applegate}

If we first consider the case where the eclipsing time variation is due to a
time-dependent quadrupole moment (classical Applegate mechanism), the expected
period variation due to a change of the gravitational
quadrupole moment can be computed from
\citep{Applegate92}
\begin{equation}
  \frac{\Delta P}{P} = -9 \frac{\Delta Q_{xx}}{Ma^2} = 2\pi\frac{\rm O-C}{P_{\rm mod}},
\end{equation}
where $\Delta P/P$ is the amplitude of the orbital period modulation, $M$ the
stellar mass, and $a$ the binary separation, ${\rm O-C}$ is the
amplitude of the observed-minus-calculated diagram and
$P_{\rm mod}$ its modulation period.
We choose V471~Tau as a reference system with which we compare our results,
The reason for this is that its main-sequence star has a mass of $M=0.93M_\odot$
and is thus structurally similar to the Sun and the current
simulations. It is also
rotating at a speed that is computationally feasible to achieve, whereas in
many other systems the main-seuqnce stars have lower masses and
rotation rates up to a hundred times the one of the Sun.
Our results are only applicate to PCEBs where the magnetically
active star has a Sun-like structure. This is because our model is constructed
to resemble such stars.
We nevertheless note that this
comparison is very approximate, as for example the Reynolds numbers of a real
stellar system cannot be numerically reproduced.

For V471~Tau, the orbital seperation is $a = 3.3R_\odot$ and there are two
contributions to the orbital period modulation that individually
result in two orbital period modulations \citep{Marchioni18}.
These are
\begin{align}
(\Delta P/P)_1 &= 8.5\times10^{-7},\\
(\Delta P/P)_2 &= 4.5\times10^{-7}.
\end{align}
The corresponding quadrupole moment variations are
\begin{align}
\Delta Q_{xx,1} &= 9.4\times10^{41}\,{\rm kg\,m}^2,\\
\Delta Q_{xx,2} &= 4.5\times10^{41}\,{\rm kg\,m}^2.
\end{align}

For the purpose of comparing with the quadrupole moment variations in
simulations, we recall here that  density fluctuations and the quadrupole
moment itself need to be scaled as explained in
\citet{Navarrete20}. They scale as
\begin{equation}
  \Delta \rho \sim \mathcal{L}_r^{2/3},
\end{equation}
where $\mathcal{L}_r$ is the ratio between the luminosities in the simulation and
the target star, that is,
\begin{equation}
  \mathcal{L}_r = \frac{\mathcal{L_{\rm sim}}}{\mathcal{L}_\star},
\end{equation}
so the quadrupole moment is accordingly scaled as
\begin{equation}
  Q_\star = \frac{1}{\mathcal{L}_r^{2/3}}Q_{\rm sim}.
\end{equation}
Details of the scaling are presented in Appendix~\ref{sec:mach}.

In Run~B the amplitude of the variation is $\Delta Q_{xx} =
1.2\times10^{40}$~kg~m$^2$ corresponding to $\Delta P/P = 1\times10^{-8}$.
By adopting a modulation period of $P_{\rm mod}~=~80$~yr, which corresponds
to the period of $Q_{xx}$, we have an observed
minus calculated value of ${\rm O-C} = 4.7$~s.
This ${\rm O-C}$ amplitude is still four and thirty times smaller than the values
reported by
\citet{Marchioni18}.
There are a few possible reasons behind this mismatch. First, we are not
including the centrifugal force and so the quadrupole moment fluctuations
are produced by the evolution of the magnetic field and the resulting
redistribution of the density rather than by the deformation of the
star like in the Applegate mechanism.
Secondly, the stars we are modeling
have convective envelopes of $30\%$ of the radius, whereas the main-sequence
star in the target system V471~Tau has a mass of $0.93M_\odot$ and therefore
a slightly more extended CZ. In a deeper CZ it is possible to perturb
the density and the angular momentum distribution in a larger portion
of the star \citep{Voelschow18}, although
we expect this contribution to be very small.
Lastly, we are imposing sphericity which is especially important at the
surface of the star. Boundary conditions that dynamically react to the
physical quantities inside of the star may allow larger variations of
the quadrupole moment, especially if the star change its shape and
size.

\subsection{Models without tidal locking}\label{subsec:comparison}

In the previous calculation of $\Delta P/P$ following \citet{Applegate92},
we made the implicit assumption of tidal
locking and that the stellar rotation axis is perpendicular to the plane of
the orbital motion. In this scenario, the $\hat{\bm x}$ axis points
towards the companion and thus it rotates together with the stellar spin.
Under those conditions, only the $Q_{xx}$ component of the gravitational
quadrupole moment contributes to the modulation of the binary period
\citep{Applegate92}. In contrast, in the scenario put forward by
\citet{Applegate89} and \citet{Lanza20}, the star is not
yet tidally locked, and its companion effectively experiences a time-varying
quadrupole moment due to the relative rotation of the magnetically active star.
This holds even if the quadrupole moment in the corotating frame of the star was
constant. This implies then that different components of $Q_{ij}$ contribute.
In the simplified \citet{Lanza20} scenario, the magnetic field is modeled as a
permanent single flux tube that lies at the equator and produces a nonaxisymmetric
density distribution and thus, a permanent nonaxisymmetric gravitational
quadrupole moment.

\begin{figure}
\includegraphics[width=\columnwidth]{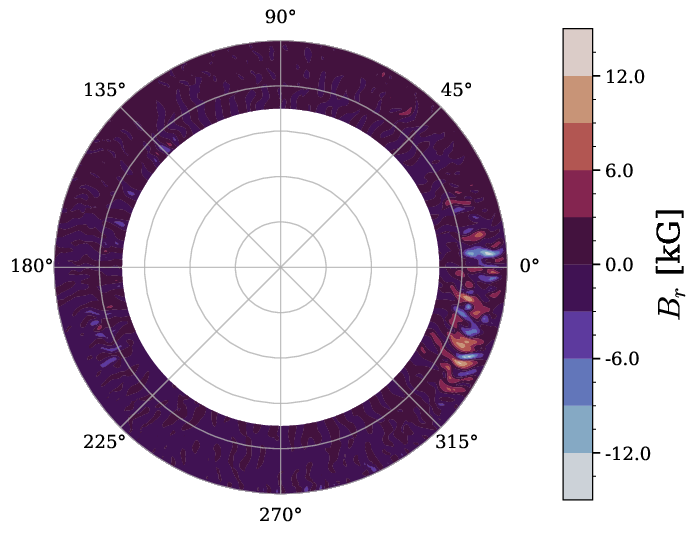}
\caption{Radial magnetic field of Run~B at the equator at
$t=155$~yr. The $\hat{\bm x}$ and
$\hat{\bm y}$ axes lie at $\phi=0^\circ$ and $\phi=90^\circ$, respectively.
The $\hat{\bm s}$ and $\hat{\bm s}'$ axes are obtained by performing
clockwise rotations.}
\label{fig:B-BrEquator}
\end{figure}

\begin{figure}
\includegraphics[width=\columnwidth]{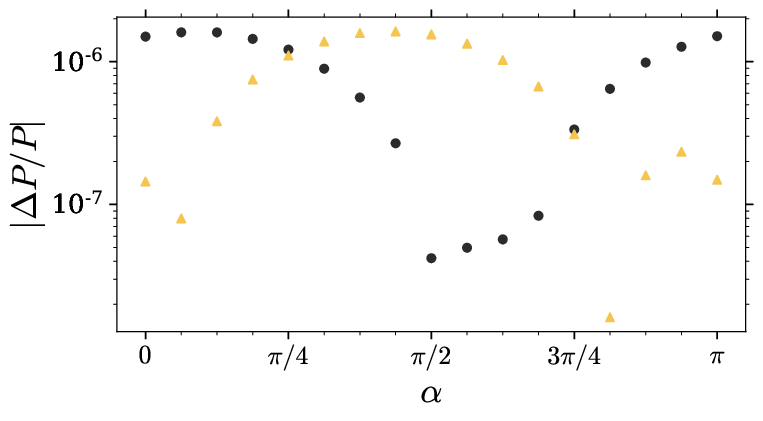}
\caption{Absolute value of $\Delta P/P$ as a function of separation
angle $\alpha$ between $\hat{\bm x}$ and $\hat{\bm s}$ for $t=110$~yr (black dots) and
$t=155$~yr (yellow triangles).}
\label{fig:B-DeltaPdivP}
\end{figure}

While there is a strong nonaxisymmetric magnetic field in our Run~B, it
is stronger at mid- and at high latitudes rather than at the
equator.
In our simulations, the choice of the $\hat{\bm x}$ and $\hat{\bm y}$ axes in the
equatorial plane along
which the moments of inertia are calculated is arbitrary,
that is, as the companion star is not being modeled. Once fixed,
we perform rotations about the $\hat{\bm z}$ axis in steps of
$\pi/16$ up to $\pi$, and then we
calculate the two moments of inertia about the rotated axes.
These axes would correspond to $\hat{\bm s}$ and $\hat{\bm s}'$ of \citet{Lanza20}.
The former is the rotated $\hat{\bm x}$ axis, and the latter is the rotated
$\hat{\bm y}$ axis. In \citet{Lanza20} $\hat{\bm s}$ is chosen to be along the
axis of symmetry of the magnetic flux tube, which is the only magnetic
structure in the CZ of the magnetically active star.
In our simulations the rotation is not unique as there is no
single radial magnetic field structure that extends from the bottom
to the surface of the CZ in our simulations that would
otherwise allow us to unequivocally choose $\hat{\bm s}$. However, a clear
radial magnetic structure at the equator is seen at
$t=155$~yr (see Fig. \ref{fig:B-BrEquator}), but magnetic fields with
different structure and strength dominate at different latitudes.
In this
configuration, the nonaxisymmetric quadrupole moment is defined as
$T=I_s-I_s'$, where $I_s$ and $I_s'$ are the moments of inertia about the
$\hat{\bm s}$ and $\hat{\bm s}'$ axes. The moment of inertia of the active
star about the spin axis is $I_p = I_{xx} + I_{yy}$. The  order of magnitude of the
period variations can then be estimated as \citep[Eq. 2 of][]{Lanza20}
\begin{equation}\label{eq:tdivp}
\frac{T}{I_p} \approx \frac{4}{3}\left(\frac{M_T}{m_S}\right)
\left(\frac{ma^2}{I_p}\right)\left(\frac{P}{P_{\rm mod}}\right)
\left|\frac{\Delta P}{P}\right|,
\end{equation}
where $M_T$ is the total mass of the binary, $m_S$ is the mass of the
companion, $m$ is the reduced mass, $P$ is the orbital period, and
$P_{\rm mod}$ is the modulation period. We take the density fields of
Run~B at $t=110$~yr and $t=155$~yr and compute the two quadrupole moments
$T$ and $I_p$. By using Eq. (\ref{eq:tdivp}) and the parameters of V471~Tau
\citep[see e.g.,][]{Hardy15, vaccaro15} we can obtain an order of
magnitude estimate of $\Delta P/P$.
Fig. \ref{fig:B-DeltaPdivP} shows the absolute value of the amplitude of the
orbital period modulation as a function of separation angle $\alpha$
between $\hat{x}$ and $\hat{s}$ for $t=110$~yr (black dots) and
$t=155$~yr (yellow triangles).
$|\Delta P/P|$ ranges between $1.5\times10^{-7}$ and $1.5\times10^{-6}$, which
contains the two contributions to the observed variations as well as their
sum \citep{Marchioni18}.
From our simulations we get a value of $I_p$ that is of the same order of
magnitude as in \cite{Lanza20}, while $T$ is about one order of magnitude
larger here. It is important to note that we have obtained $T$ based on a detailed
3D magneto-hydrodynamical simulation, while \citet{Lanza20} simply calculated which
$T$ would be required to explain the observed ETVs.

In general, the gravitational potential felt by the companion can be written as
\citep{Applegate92, Lanza20}
\begin{equation}
    \Phi = -\frac{GM}{r} - \frac{3G}{2r^3}\sum_{i,j}\frac{Q_{ij}x_ix_j}{r^2},
\end{equation}
where $G$ is the gravitational constant, $M$ is the mass of the active star,
$r$ is the distance between the center of the active star and the companion,
$Q_{ij}$ is the quadrupole moment tensor, and ${\bm x}$ refers to Cartesian coordinates.
Writing out the summation explicitly and expressing $x_i$ and $x_j$ in a
spherical coordinate system $(r,\theta',\phi')$ with its origin coinciding with
the center of the star, we arrive at
\begin{align}\label{eq:gravpot}
    \Phi =& -\frac{GM}{r} - \frac{3G}{2r^3}
        \biggl\{ Q_{xx}\sin^2\theta'\cos^2\phi' \nonumber \\
        & + Q_{yy}\sin^2\theta'\sin^2\phi' \nonumber \\
        & + Q_{zz}\cos^2\theta' \nonumber \\
        & + 2 \biggl(
                Q_{xy}\sin^2\theta'\cos\phi'\sin\phi' \nonumber \\
                & + Q_{xz}\sin\theta'\cos\theta'\cos\phi' \nonumber \\
                & + Q_{yz}\sin\theta'\cos\theta'\sin\phi'\biggr)\biggr\}.
\end{align}
The case of $\theta' = \pi/2$ and $\phi' = 0$ is analogous to the assumptions
that the rotation axis is perpendicular to the plane of the orbit and
that the
orbital motion is tidally locked, respectively. By assuming only
the former, Eq. (\ref{eq:gravpot}) is reduced to
\begin{align}\label{eq:gravpotred}
    \Phi =& -\frac{GM}{r} - \frac{3G}{2r^3}\biggl(Q_{xx}\cos^2\phi' \nonumber \\
         & + Q_{yy}\sin^2\phi' +2Q_{xy}\cos\phi'\sin\phi'\biggr).
\end{align}
Here the effects of deviations from tidal locking can be
modeled by making $\phi'$ time-dependent. There are two
alternatives, namely
\begin{align}
    \phi'_1 &= \alpha \cos (\omega t), \\
    \phi'_2 &= \omega t.
\end{align}
In the former case, the companion is seen in the frame of reference of the rotating
star as oscillating in the orbital plane with amplitude $\alpha$
and angular velocity $\omega$. In that case, $\phi'_1$ corresponds to the
analogous of the libration model. The latter expression
for $\phi'_2$ corresponds to the
circulation model presented by \cite{Lanza20}. This
introduces two further contributions to the binary period variation
that come from $Q_{yy}$ and $Q_{xy}$ (see Eq. \ref{eq:gravpotred}).
In Run~B $Q_{yy}$ is, on average, the same as $Q_{xx}$. Meanwhile,
$Q_{xy}$ is $10^2...10^3$ times smaller so it can be neglected. Thus,
\begin{equation}
    \Phi = -\frac{GM}{r} - \frac{3G}{2r^3}\biggl(Q_{xx}\cos^2\phi' + Q_{yy}\sin^2\phi' \biggl).
\end{equation}

In contrast to previous studies, we can directly calculate
each component of the gravitational quadrupole moment from
our simulations. In this case it is advantageous to use Eqs. (\ref{eq:gravpot})
and (\ref{eq:gravpotred}) rather than taking the limit of $\phi'=0$.
However, we would need to use new expressions to derive $\Delta P/P$ considering
the libration and circulation models.
Alternatively, it is also possible to try different values of $\alpha$
and $\omega$, and then directly solve Eq. (\ref{eq:gravpot}) in a
two-body simulation, which is however beyond the scope of the presented study.
The influence of differences between $Q_{xx}$ and $Q_{yy}$ can be studied
with N-body simulations by prescribing their time evolution and varying their
amplitudes. It would be interesting to derive the parameters that can reproduce the
observations and to compare them with our simulations.

\section{Conclusions}\label{sect:conclusions}

We have presented three MHD simulations of stellar convection
with different rotation rates, and studied the gravitational quadrupole
moment and its connection to dynamo-generated magnetic
fields. The analysis is based on a spherical harmonic decomposition of density
and magnetic fields. Our results for Run~B ($P=1.2$~days) show that a hemispheric
dynamo mode
can be an important ingredient for the eclipsing time variations in close
binaries. This hemispheric dynamo produces equatorially asymmetric
density variations and changes the moment of inertia along the
rotation axis. The hemispheric activity migrates seemingly
periodically between hemispheres and modulates
the gravitational quadrupole moment. Furthermore, nonaxisymmetric magnetic
fields modulate the other two diagonal components
of the inertia tensor, adding a further modulation of $Q$. We also expect to
have a further modulation of $Q$ that comes from the
centrifugal force which will be included in a future work as it is the
responsible for the angular momentum redistribution in the Applegate mechanism
\citep[][]{Applegate92}. Linear correlation analysis confirms the
role of the magnetic field in changing the quadrupole moment
via density variations (Table \ref{tab:corr}) and the scatter plot between
magnetic energy and quadrupole moment shows that large quadrupole moments are
related to increased magnetic energy (see Fig. \ref{fig:QxxEmag}).

When our results are interpreted in the context of the classical Applegate
mechanism, that is the
star is tidally locked, then only the $Q_{xx}$ component of the quadrupole
moment contributes to the period variations. In this scenario, we obtain
orbital period modulations between one and two orders of magnitude smaller
than observed in the target system V471~Tau \citep[][]{Marchioni18}.
We emphasize that our results here should be taken with caution. We model the
CZ of a Sun-like star while the CZ extends inward for less massive stars which
are more common among PCEBs. It is
yet to be investigated if large enough quadrupole moments are found in
magnetohydrodynamical simulations of fully convective stars.

In the context of the  models by \citet{Applegate89} and \citet{Lanza20}, the order
of magnitude estimate of the amplitude of the period modulation is
$10^{-6}\hdots10^{-7}$. This range encompasses the two observed contributions
to the ${\rm O-C}$ diagram, as well as their combined effect.
The observed period variations could be a combination of both, namely, both the
axi- and nonaxisymmetric quadrupole moments contribute to them.
The implication of the first
interpretation is that there must be a hemispheric dynamo with an alternating
active hemisphere in order to modulate $Q_{xx}$ as seen in our simulations.
The second interpretation implies that the star is not tidally locked and that
there is a nonaxisymmetric magnetic field in the CZ of the magnetically active
star.
We emphasize, however, given the caveats of the model such as
imposed spherical symmetry, the coincidence in the order of magnitude between the
ETVs, and in our model must be taken with
caution. More importantly,
relaxing the assumption of tidal locking leads to period variations
that are between one and two orders of magnitude larger than in the
tidal-locking scenario.

Observational studies suggest that both scenarios discussed, namely asymmetric
magnetic fields and nontidally locked stars, are plausible.
Firstly, a recent study by \cite{Klein21}
reported the reconstruction of the surface magnetic field of Proxima Centauri
using Zeeman-Doppler imaging (ZDI). They found that the magnetic
field is mainly
poloidal with a dominant feature that is tilted at $51^\circ$ to the rotation
axis (see their Figure 3) with a strength of $135$~G, that is a field distribution
that is asymmetric with respect to the equator. This is a rather weak
field so density fluctuations should be smaller than what we find in our
simulations. However, Proxima Centauri is a slowly rotating M5.5 fully-convective
star. The magnetic field strength of fully-convective stars increases with
rotation until a saturation regime is reached, as measured by X-ray emission
\citep[see e.g.,][]{Wright16}, so density variations in magnetically active
components of PCEBs are expected to be larger due to the increased magnetic
field strength. This might also be the case for
more evolved partially convective stars as a similar scaling property was
recently found \citep{Lehtinen20}. Studying the differences of stellar spots
during a minima and maxima of of $O-C$ diagrams in PCEBs will provide direct
evidence of the connection between the underlying dynamo and the orbital
period variations.
Secondly, the determination of tidal synchronization is equally important,
as a deviation from synchronization results in a more complex relation
between the gravitational quadrupole moment and eclipsing time variations
and potentially larger binary period variations
\citep[see][and also Sect. \ref{subsec:comparison} of this paper]{Lanza20}.
\cite{Lurie17} studied tidal synchronization of F, G, and K stars in
short-period binaries. The authors find 21 eclipsing binaries that are
not synchronized and argue that this could be explained either because
they are young or have a complex dynamical history. Considering the
dynamical evolution of PCEBs, where the
secondary star is engulfed by the companion and spirals inwards toward
the core of the more massive star \citep{Paczynski76}, it is
conceivable that they fall in this category.

The determination of the degree of synchronization in post common envelope
binaries would be beneficial to further improve the understanding of the
ETVs. The surface magnetic field distribution would be
as equally important because such nonaxisymmetric fields produce larger
quadrupole moments. Furthermore, there is also unexplored grounds in the
simulations, such as the impact of the centrifugal force. Numerical models,
specially for fully convective stars such as those in \citet{Kapyla21},
that allow more freedom on the surface and near-surface layers of stars are
desired as changes in the oblateness of the star can be captured.


\section*{Acknowledgements}

FHN acknowledges financial support from the DAAD (Deutscher Akademischer
Austauschdienst; code 91723643) for his doctoral studies.
PJK acknowledges the financial support from the DFG (Deutsche
Forschungsgemeinschaft) Heisenberg programme grant No.\ KA 4825/2-1.
DRGS thanks for funding via Fondecyt regular (project code 1201280), ANID Programa
de Astronomia Fondo Quimal QUIMAL170001 and the BASAL
Centro de Excelencia en Astrofisica y Tecnologias Afines (CATA) grant
PFB-06/2007. RB acknowledges support by the DFG under Germany's Excellence Strategy
-- EXC 2121 "Quantum Universe" -- 390833306. RB is also thankful for funding by the
DFG through the projects No. BA 3706/14-1, No. BA 3706/15-1, No. BA 3706/17-1, and
No. BA 3706/18.
The simulations were run on the Leftraru/Guacolda supercomputing
cluster hosted by the NLHPC (ECM-02), the Kultrun cluster hosted at the
Departamento de Astronom\'ia, Universidad de Concepci\'on, and on HLRN-IV under
project grant hhp00052.


\bibliographystyle{aa}
\bibliography{non-axisB-Qxx}

\begin{thebibliography}{33}
\expandafter\ifx\csname natexlab\endcsname\relax\def\natexlab#1{#1}\fi

\bibitem[{{Applegate}(1989)}]{Applegate89}
{Applegate}, J.~H. 1989, \apj, 337, 865

\bibitem[{{Applegate}(1992)}]{Applegate92}
{Applegate}, J.~H. 1992, \apj, 385, 621

\bibitem[{{Brinkworth} {et~al.}(2006){Brinkworth}, {Marsh}, {Dhillon}, \&
  {Knigge}}]{Brinkworth06}
{Brinkworth}, C.~S., {Marsh}, T.~R., {Dhillon}, V.~S., \& {Knigge}, C. 2006,
  \mnras, 365, 287

\bibitem[{{Brown} {et~al.}(2020){Brown}, {Oishi}, {Vasil}, {Lecoanet}, \&
  {Burns}}]{Brown20}
{Brown}, B.~P., {Oishi}, J.~S., {Vasil}, G.~M., {Lecoanet}, D., \& {Burns},
  K.~J. 2020, \apjl, 902, L3

\bibitem[{{Busse}(2002)}]{Busse02}
{Busse}, F.~H. 2002, Physics of Fluids, 14, 1301

\bibitem[{{Cole} {et~al.}(2014){Cole}, {K{\"a}pyl{\"a}}, {Mantere}, \&
  {Brandenburg}}]{Cole14}
{Cole}, E., {K{\"a}pyl{\"a}}, P.~J., {Mantere}, M.~J., \& {Brandenburg}, A.
  2014, \apjl, 780, L22

\bibitem[{{Gastine} {et~al.}(2014){Gastine}, {Yadav}, {Morin}, {Reiners}, \&
  {Wicht}}]{Gastine14}
{Gastine}, T., {Yadav}, R.~K., {Morin}, J., {Reiners}, A., \& {Wicht}, J. 2014,
  \mnras, 438, L76

\bibitem[{{Grote} \& {Busse}(2000)}]{Grote&Busse00}
{Grote}, E. \& {Busse}, F.~H. 2000, \pre, 62, 4457

\bibitem[{{Hardy} {et~al.}(2015){Hardy}, {Schreiber}, {Parsons}, {Caceres},
  {Retamales}, {Wahhaj}, {Mawet}, {Canovas}, {Cieza}, {Marsh}, {Bours},
  {Dhillon}, \& {Bayo}}]{Hardy15}
{Hardy}, A., {Schreiber}, M.~R., {Parsons}, S.~G., {et~al.} 2015, \apjl, 800,
  L24

\bibitem[{{K{\"a}pyl{\"a}}(2021)}]{Kapyla21}
{K{\"a}pyl{\"a}}, P.~J. 2021, \aap, 651, A66

\bibitem[{{K{\"a}pyl{\"a}} {et~al.}(2014){K{\"a}pyl{\"a}}, {K{\"a}pyl{\"a}}, \&
  {Brandenburg}}]{Kapyla14}
{K{\"a}pyl{\"a}}, P.~J., {K{\"a}pyl{\"a}}, M.~J., \& {Brandenburg}, A. 2014,
  \aap, 570, A43

\bibitem[{{K{\"a}pyl{\"a}} {et~al.}(2010){K{\"a}pyl{\"a}}, {Korpi},
  {Brandenburg}, {Mitra}, \& {Tavakol}}]{Kapyla10}
{K{\"a}pyl{\"a}}, P.~J., {Korpi}, M.~J., {Brandenburg}, A., {Mitra}, D., \&
  {Tavakol}, R. 2010, Astronomische Nachrichten, 331, 73

\bibitem[{{K{\"a}pyl{\"a}} {et~al.}(2013){K{\"a}pyl{\"a}}, {Mantere}, {Cole},
  {Warnecke}, \& {Brandenburg}}]{Kapyla13}
{K{\"a}pyl{\"a}}, P.~J., {Mantere}, M.~J., {Cole}, E., {Warnecke}, J., \&
  {Brandenburg}, A. 2013, \apj, 778, 41

\bibitem[{{Klein} {et~al.}(2021){Klein}, {Donati}, {H{\'e}brard}, {Zaire},
  {Folsom}, {Morin}, {Delfosse}, \& {Bonfils}}]{Klein21}
{Klein}, B., {Donati}, J.-F., {H{\'e}brard}, {\'E}.~M., {et~al.} 2021, \mnras,
  500, 1844

\bibitem[{{Krause} \& {R\"adler}(1980)}]{Krause80}
{Krause}, F. \& {R\"adler}, K.-H. 1980, {Mean-field magnetohydrodynamics and
  dynamo theory} (Oxford, Pergamon Press, Ltd., 1980.~271 p.)

\bibitem[{{Lanza}(2020)}]{Lanza20}
{Lanza}, A.~F. 2020, \mnras, 491, 1820

\bibitem[{{Lanza} {et~al.}(1998){Lanza}, {Rodono}, \& {Rosner}}]{Lanza98}
{Lanza}, A.~F., {Rodono}, M., \& {Rosner}, R. 1998, \mnras, 296, 893

\bibitem[{{Lehtinen} {et~al.}(2020){Lehtinen}, {Spada}, {K{\"a}pyl{\"a}},
  {Olspert}, \& {K{\"a}pyl{\"a}}}]{Lehtinen20}
{Lehtinen}, J.~J., {Spada}, F., {K{\"a}pyl{\"a}}, M.~J., {Olspert}, N., \&
  {K{\"a}pyl{\"a}}, P.~J. 2020, Nature Astronomy, 4, 658

\bibitem[{{Lurie} {et~al.}(2017){Lurie}, {Vyhmeister}, {Hawley}, {Adilia},
  {Chen}, {Davenport}, {Juri{\'c}}, {Puig-Holzman}, \&
  {Weisenburger}}]{Lurie17}
{Lurie}, J.~C., {Vyhmeister}, K., {Hawley}, S.~L., {et~al.} 2017, \aj, 154, 250

\bibitem[{{Marchioni} {et~al.}(2018){Marchioni}, {Guinan}, {Engle}, {Dowling
  Jones}, {Michail}, {Werner}, \& {Ribas}}]{Marchioni18}
{Marchioni}, L., {Guinan}, E.~F., {Engle}, S.~G., {et~al.} 2018, Research Notes
  of the American Astronomical Society, 2, 179

\bibitem[{{Navarrete} {et~al.}(2020){Navarrete}, {Schleicher},
  {K{\"a}pyl{\"a}}, {Schober}, {V{\"o}lschow}, \& {Mennickent}}]{Navarrete20}
{Navarrete}, F.~H., {Schleicher}, D. R.~G., {K{\"a}pyl{\"a}}, P.~J., {et~al.}
  2020, \mnras, 491, 1043

\bibitem[{{Navarrete} {et~al.}(2021){Navarrete}, {Schleicher},
  {K{\"a}pyl{\"a}}, {Schober}, {V{\"o}lschow}, \&
  {Mennickent}}]{2021MNRAS.504.1676N}
{Navarrete}, F.~H., {Schleicher}, D. R.~G., {K{\"a}pyl{\"a}}, P.~J., {et~al.}
  2021, \mnras, 504, 1676

\bibitem[{{Navarrete} {et~al.}(2018){Navarrete}, {Schleicher}, {Zamponi
  Fuentealba}, \& {V{\"o}lschow}}]{Navarrete18}
{Navarrete}, F.~H., {Schleicher}, D.~R.~G., {Zamponi Fuentealba}, J., \&
  {V{\"o}lschow}, M. 2018, \aap, 615, A81

\bibitem[{{Paczynski}(1976)}]{Paczynski76}
{Paczynski}, B. 1976, in IAU Symposium, Vol.~73, Structure and Evolution of
  Close Binary Systems, ed. P.~{Eggleton}, S.~{Mitton}, \& J.~{Whelan}, 75

\bibitem[{{Pencil Code Collaboration} {et~al.}(2021){Pencil Code
  Collaboration}, {Brandenburg}, {Johansen}, {Bourdin}, {Dobler}, {Lyra},
  {Rheinhardt}, {Bingert}, {Haugen}, {Mee}, {Gent}, {Babkovskaia}, {Yang},
  {Heinemann}, {Dintrans}, {Mitra}, {Candelaresi}, {Warnecke},
  {K{\"a}pyl{\"a}}, {Schreiber}, {Chatterjee}, {K{\"a}pyl{\"a}}, {Li},
  {Kr{\"u}ger}, {Aarnes}, {Sarson}, {Oishi}, {Schober}, {Plasson}, {Sandin},
  {Karchniwy}, {Rodrigues}, {Hubbard}, {Guerrero}, {Snodin}, {Losada},
  {Pekkil{\"a}}, \& {Qian}}]{2021JOSS....6.2807P}
{Pencil Code Collaboration}, {Brandenburg}, A., {Johansen}, A., {et~al.} 2021,
  The Journal of Open Source Software, 6, 2807

\bibitem[{{Vaccaro} {et~al.}(2015){Vaccaro}, {Wilson}, {Van Hamme}, \&
  {Terrell}}]{vaccaro15}
{Vaccaro}, T.~R., {Wilson}, R.~E., {Van Hamme}, W., \& {Terrell}, D. 2015,
  \apj, 810, 157

\bibitem[{{Vanderbosch} {et~al.}(2017){Vanderbosch}, {Clemens}, {Dunlap}, \&
  {Winget}}]{Vanderbosch17}
{Vanderbosch}, Z.~P., {Clemens}, J.~C., {Dunlap}, B.~H., \& {Winget}, D.~E.
  2017, in Astronomical Society of the Pacific Conference Series, Vol. 509,
  20th European White Dwarf Workshop, ed. P.-E. {Tremblay}, B.~{Gaensicke}, \&
  T.~{Marsh}, 571--574

\bibitem[{{Viviani} {et~al.}(2018){Viviani}, {Warnecke}, {K{\"a}pyl{\"a}},
  {K{\"a}pyl{\"a}}, {Olspert}, {Cole-Kodikara}, {Lehtinen}, \&
  {Brandenburg}}]{Viviani18}
{Viviani}, M., {Warnecke}, J., {K{\"a}pyl{\"a}}, M.~J., {et~al.} 2018, \aap,
  616, A160

\bibitem[{{V{\"o}lschow} {et~al.}(2014){V{\"o}lschow}, {Banerjee}, \&
  {Hessman}}]{Voelschow14}
{V{\"o}lschow}, M., {Banerjee}, R., \& {Hessman}, F.~V. 2014, \aap, 562, A19

\bibitem[{{V{\"o}lschow} {et~al.}(2018){V{\"o}lschow}, {Schleicher},
  {Banerjee}, \& {Schmitt}}]{Voelschow18}
{V{\"o}lschow}, M., {Schleicher}, D.~R.~G., {Banerjee}, R., \& {Schmitt},
  J.~H.~M.~M. 2018, \aap, 620, A42

\bibitem[{{V{\"o}lschow} {et~al.}(2016){V{\"o}lschow}, {Schleicher},
  {Perdelwitz}, \& {Banerjee}}]{Voelschow16}
{V{\"o}lschow}, M., {Schleicher}, D.~R.~G., {Perdelwitz}, V., \& {Banerjee}, R.
  2016, \aap, 587, A34

\bibitem[{{Wright} \& {Drake}(2016)}]{Wright16}
{Wright}, N.~J. \& {Drake}, J.~J. 2016, \nat, 535, 526

\bibitem[{{Zorotovic} \& {Schreiber}(2013)}]{zorotovic13}
{Zorotovic}, M. \& {Schreiber}, M.~R. 2013, \aap, 549, A95

\end{thebibliography}


\newpage
\appendix

\section{Scaling of the quadrupole moment with Mach number}\label{sec:mach}

In \citet{Navarrete20} the scaling of the quadrupole moment was assumed
to be
\begin{equation}
  Q_{xx,{\rm phys}} \propto {\cal L}_r^{-2/3} Q_{xx,{\rm code}},
\end{equation}
where $Q_{xx,{\rm phys}}$ is the physical quadrupole moment,
and ${\cal L}_r$\footnote{${\cal L}_r$ corresponds to $\mathfrak{F}_r$
  of \citet{Navarrete20}.} is the ratio of simulation to solar
luminosities at the bottom of the CZ, and
$Q_{xx,{\rm code}}$ is the quadrupole moment in code units.
Equivalently,
\begin{equation}
    Q_{xx,{\rm phys}} \propto {\rm Ma}^{2} Q_{xx,{\rm code}},
\end{equation}
where Ma is the Mach number.
We test this scaling with two sets of simulations. Set~L uses the same parameters
as in Run~B but we omit magnetic fields and rotation. Set~M includes both
rotation and magnetic fields.
In particular, Run~L2 has the same input parameters as Run~B such as
${\cal L}_r$, but without rotation and magnetic fields.
Relevant quantities are shown in Table \ref{tab:parameters}.
Run~L2 produces a
maximum variation of the quadrupole moment
$\Delta Q_{xx} = 6.8\times10^{38}\,{\rm kg\,m}^2$, which is about 18 times
smaller than in Run~B. These variations develop on a timescale of 5
years and remain below the aforementioned level after that.

\begin{figure}
\includegraphics[width=\columnwidth]{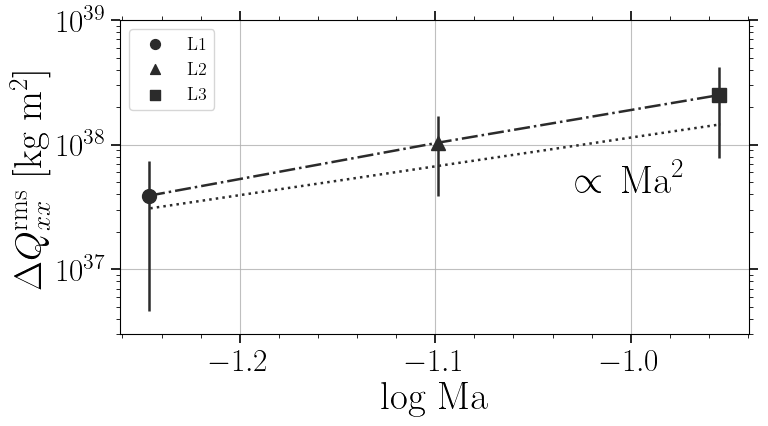}
\caption{Root-mean-squared quadrupole moment fluctuations as a function of
Mach number for Set L (without rotation and magnetic fields). The dotted line
is proportional to the Mach number squared and the dash-dotted line joins the
data points.}
\label{fig:MachL}
\end{figure}

\begin{figure}
\includegraphics[width=\columnwidth]{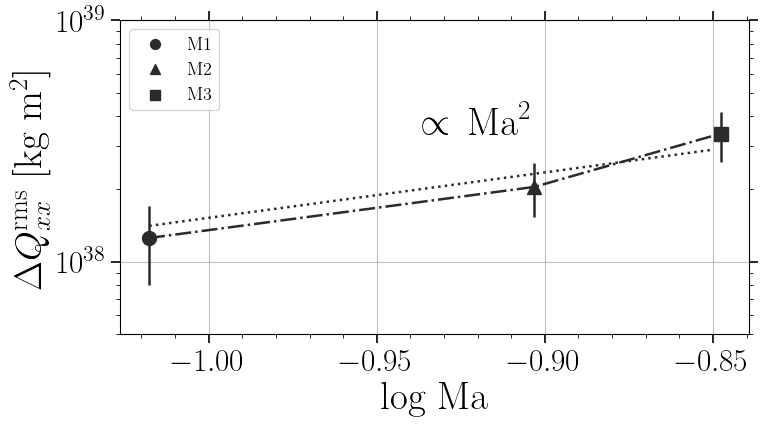}
\caption{Root-mean-squared quadrupole moment fluctuations as a function of
Mach number for Set M (with rotation and magnetic fields). The dotted line
is proportional to the Mach number squared and the dash-dotted line joins the
data points.}
\label{fig:MachM}
\end{figure}

The rms value of the quadrupole moment fluctuations as
a function of Mach number is shown in Fig.
\ref{fig:MachL} for Set~L and in Fig. \ref{fig:MachM} for Set~M.
The results are in reasonable agreement with the theoretical scaling, which
is indicated by the dotted line in each plot.

\begin{table}
\caption{Parameters of sets L and M.}
\label{tab:scaling1}
\begin{tabular}{l | c c c}
\hline
Run & $\mathfrak{F}_r$ & Ma & $\Delta Q_{xx}^{\rm rms}$ \\
\hline
L1 & $2.74\times10^5$ & $5.67\times10^{-2}$ & $1.07\times10^{-5}$ \\ 
L2 & $8.07\times10^5$ & $7.97\times10^{-2}$ & $2.85\times10^{-5}$ \\
L3 & $2.34\times10^6$ & $1.11\times10^{-1}$ & $6.87\times10^{-5}$ \\
M1 & $2.12\times10^5$ & $9.60\times10^{-2}$ & $3.44\times10^{-5}$ \\
M2 & $6.37\times10^5$ & $1.25\times10^{-1}$ & $5.60\times10^{-5}$ \\
M3 & $2.12\times10^6$ & $1.42\times10^{-1}$ & $9.30\times10^{-5}$ \\
\hline
\end{tabular}
\end{table}

\section{Figures of the decomposed fields}\label{sec:decomposed}
\begin{figure*}
\includegraphics[width=0.33\linewidth]{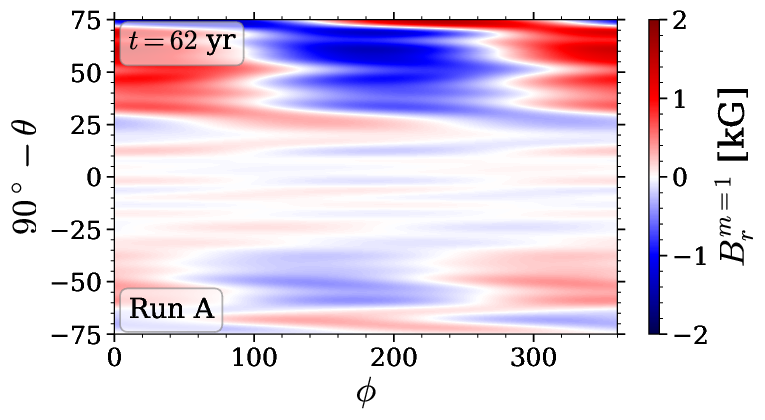}
\includegraphics[width=0.33\linewidth]{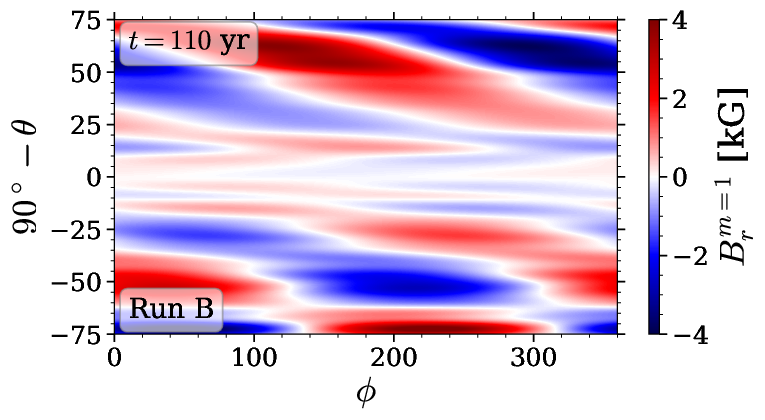}
\includegraphics[width=0.33\linewidth]{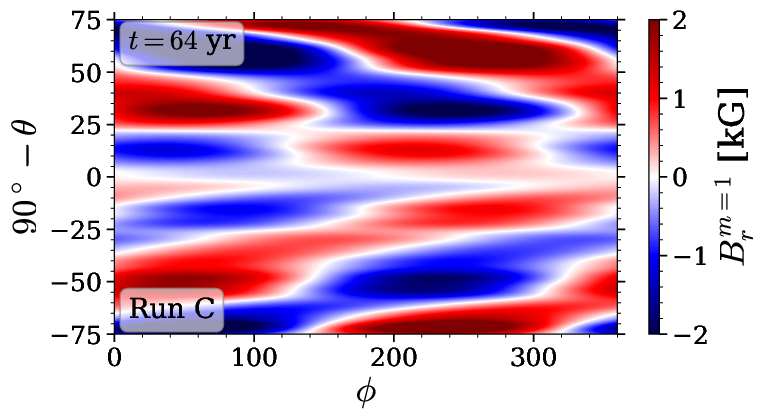}
\includegraphics[width=0.33\linewidth]{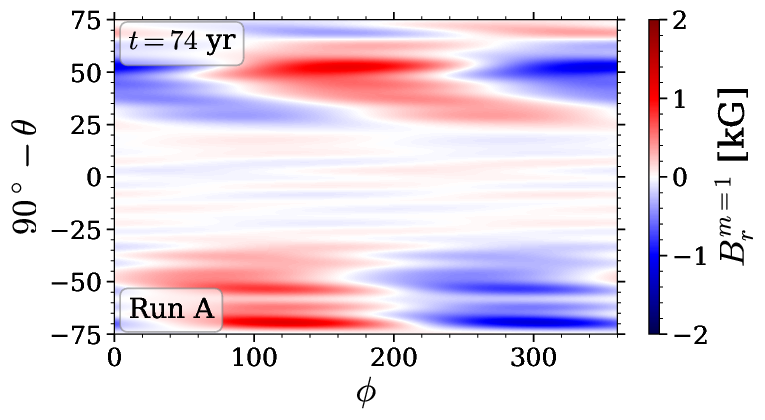}
\includegraphics[width=0.33\linewidth]{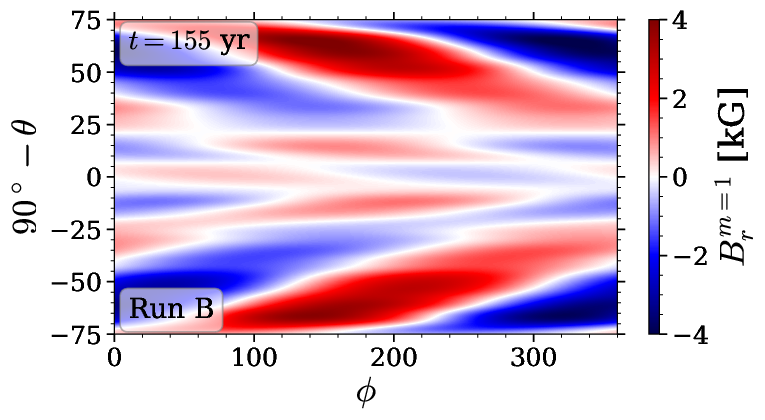}
\includegraphics[width=0.33\linewidth]{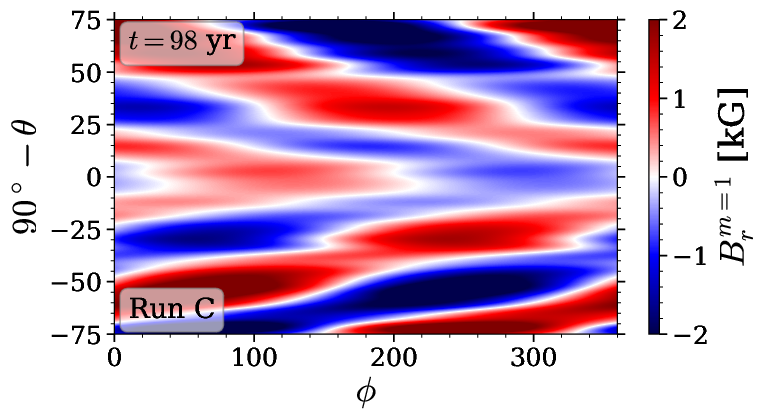}
\caption{First nonaxisymmetric mode of the radial magnetic field ($B_r^{m=1}$)
at $r=0.98R$ for each run.}
\label{fig:Brm1}
\end{figure*}

\begin{figure*}
\includegraphics[width=0.33\linewidth]{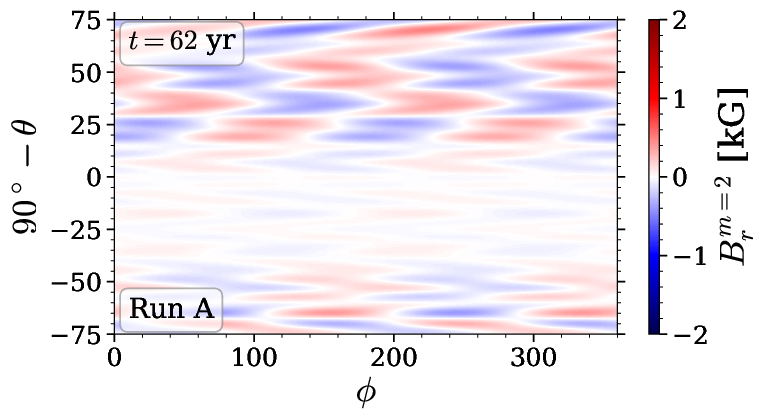}
\includegraphics[width=0.33\linewidth]{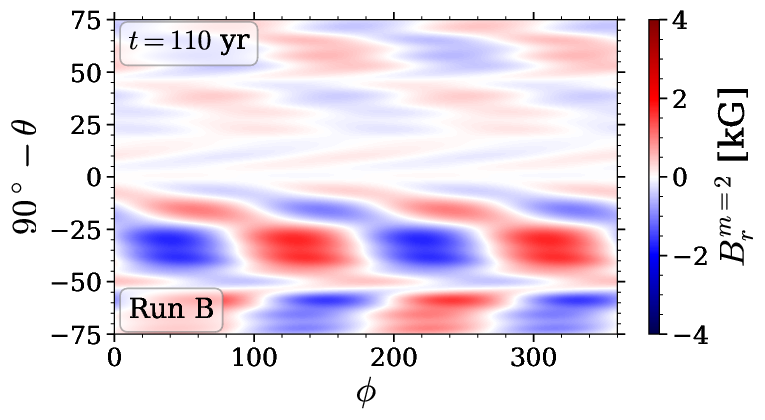}
\includegraphics[width=0.33\linewidth]{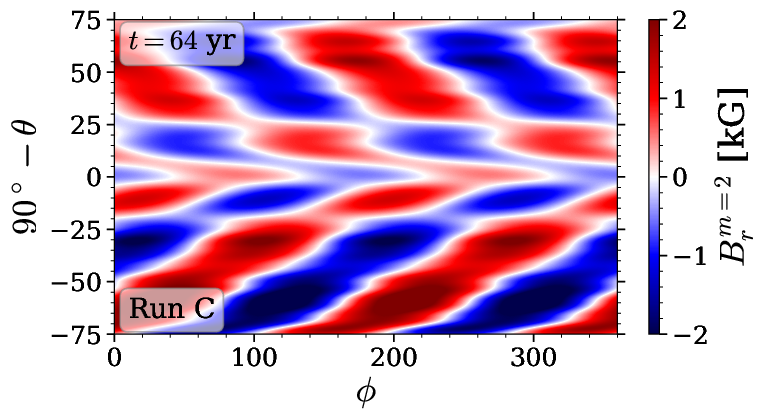}
\includegraphics[width=0.33\linewidth]{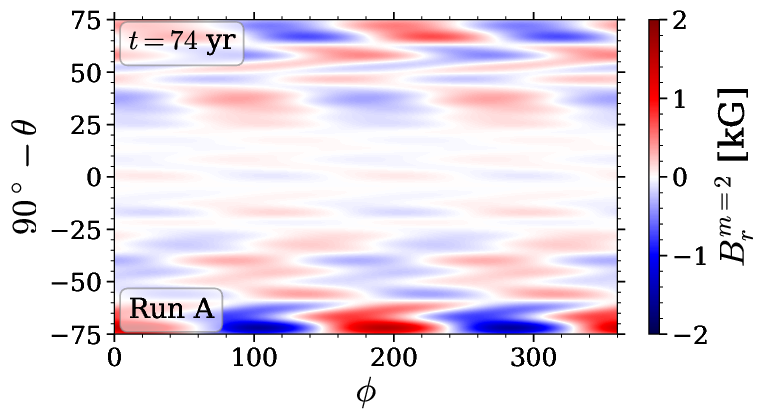}
\includegraphics[width=0.33\linewidth]{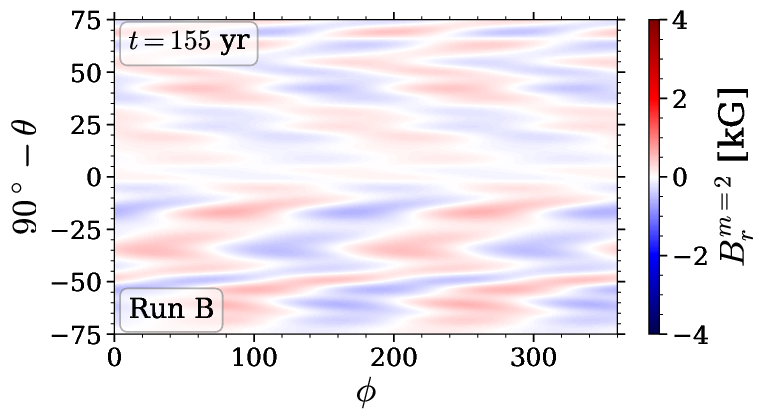}
\includegraphics[width=0.33\linewidth]{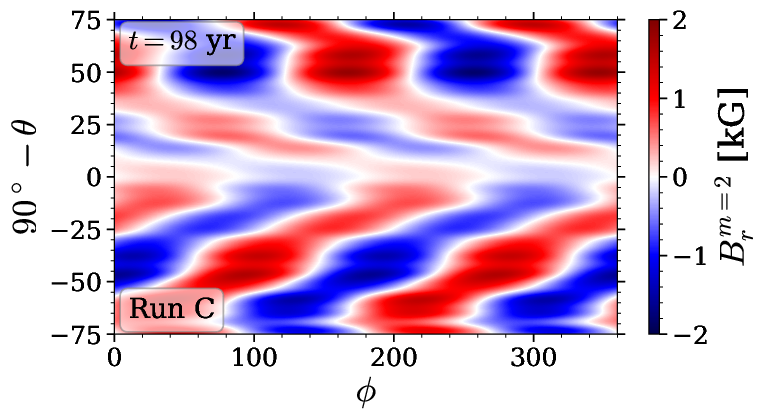}
\caption{Second nonaxisymmetric mode of the radial magnetic field ($B_r^{m=2}$)
at $r=0.98R$ for each run.}
\label{fig:Brm2}
\end{figure*}

\begin{figure*}
\includegraphics[width=0.33\linewidth]{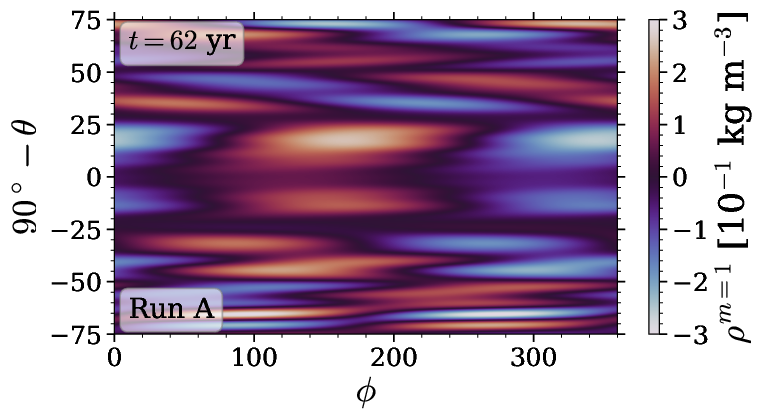}
\includegraphics[width=0.33\linewidth]{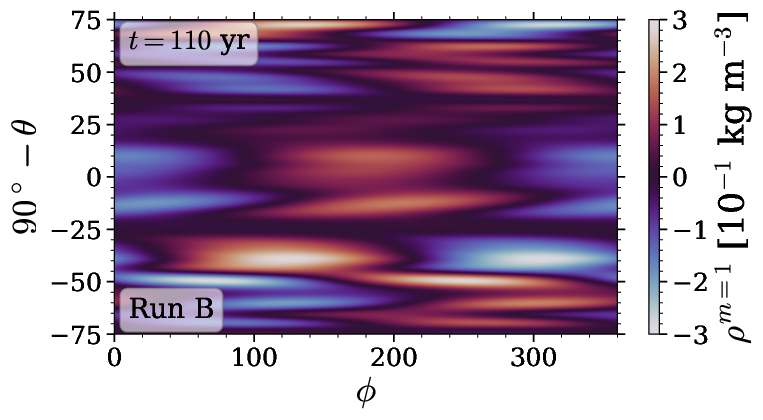}
\includegraphics[width=0.33\linewidth]{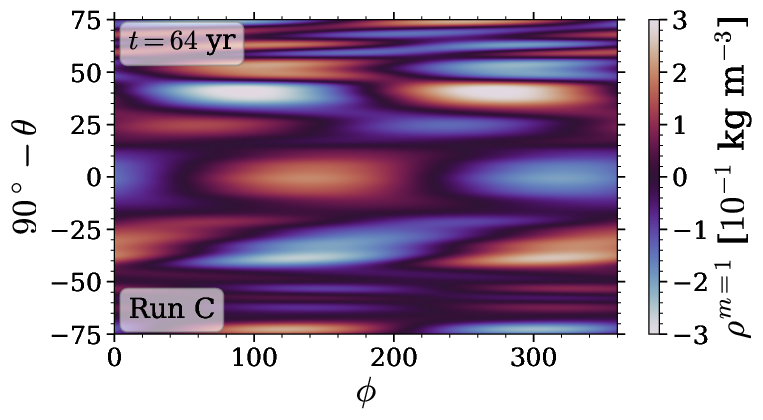}
\includegraphics[width=0.33\linewidth]{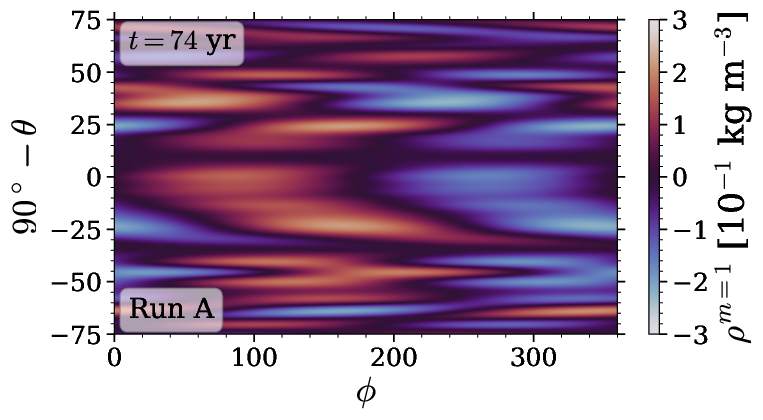}
\includegraphics[width=0.33\linewidth]{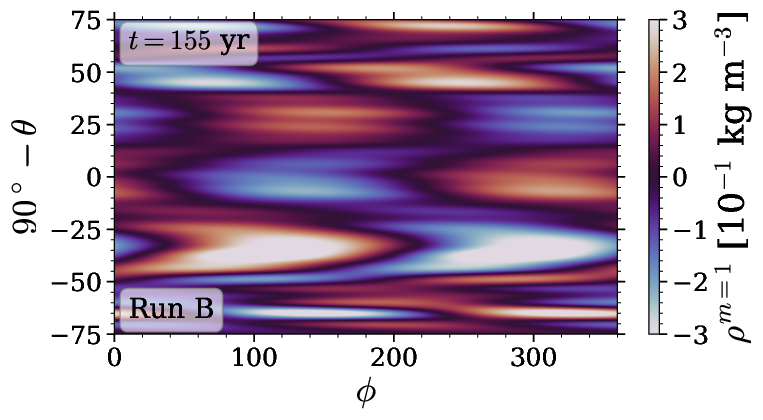}
\includegraphics[width=0.33\linewidth]{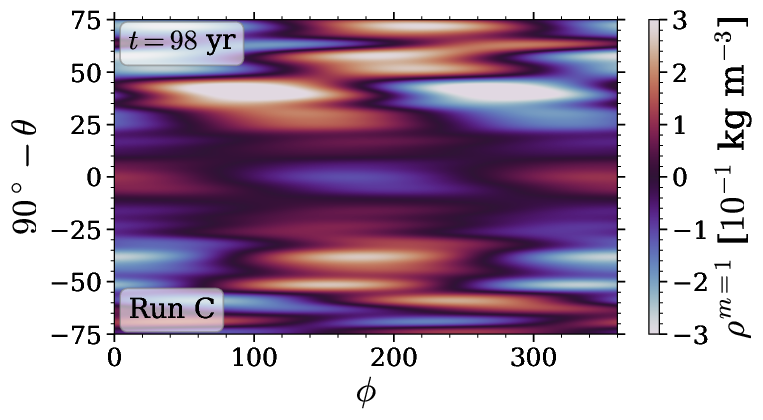}
\caption{First nonaxisymmetric mode of density at $r=0.98R$ for each Run.}
\label{fig:rhom1}
\end{figure*}
\begin{figure*}
\includegraphics[width=0.33\linewidth]{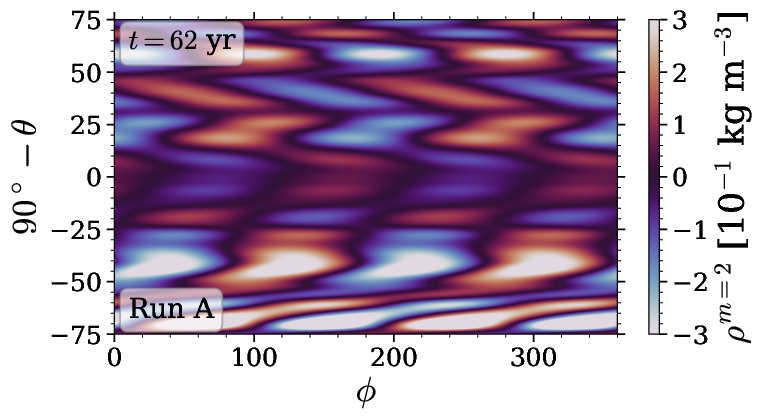}
\includegraphics[width=0.33\linewidth]{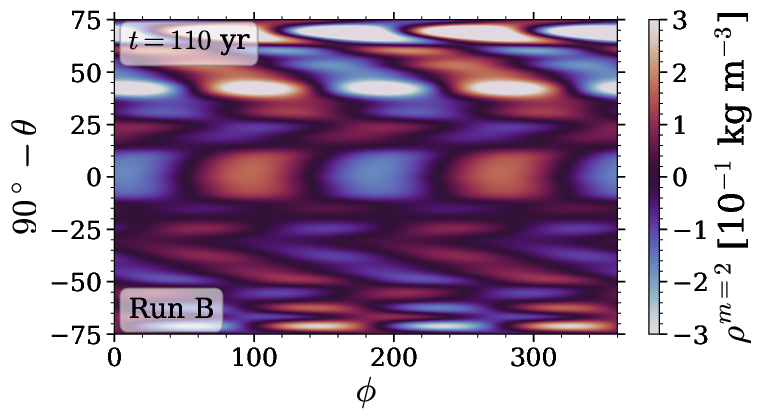}
\includegraphics[width=0.33\linewidth]{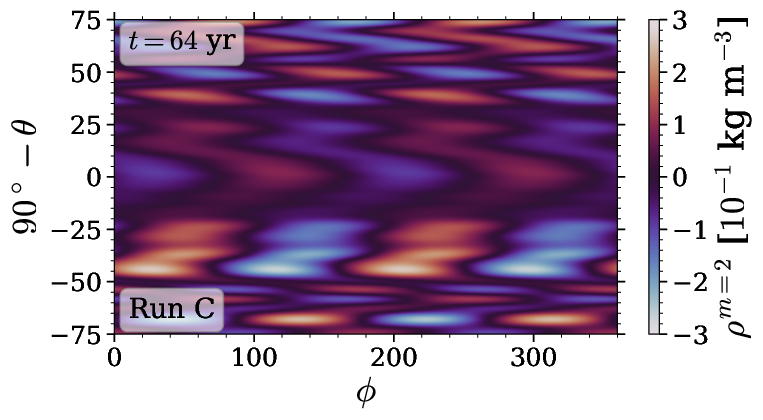}
\includegraphics[width=0.33\linewidth]{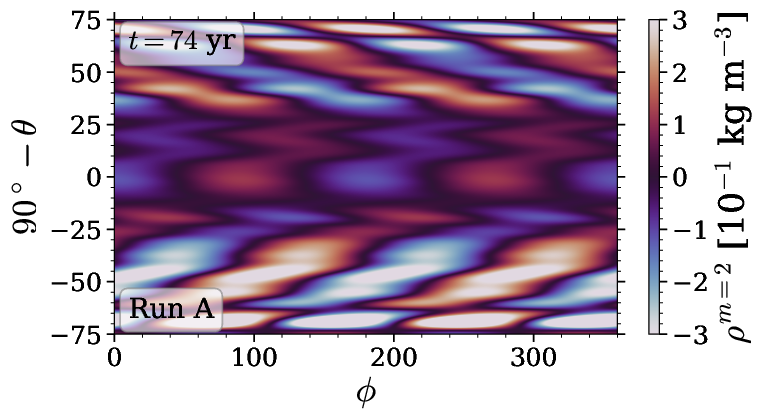}
\includegraphics[width=0.33\linewidth]{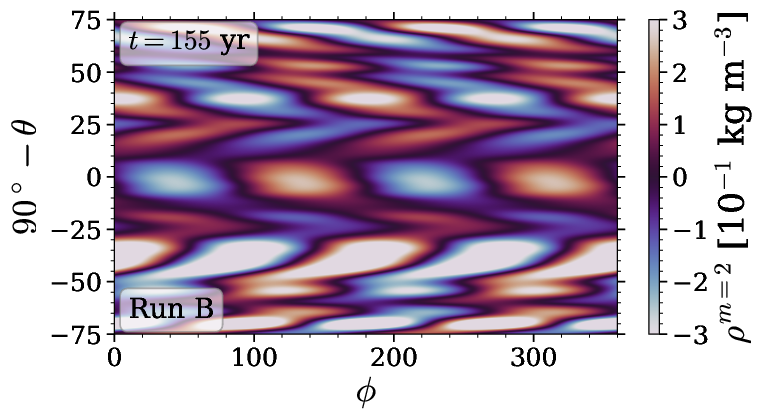}
\includegraphics[width=0.33\linewidth]{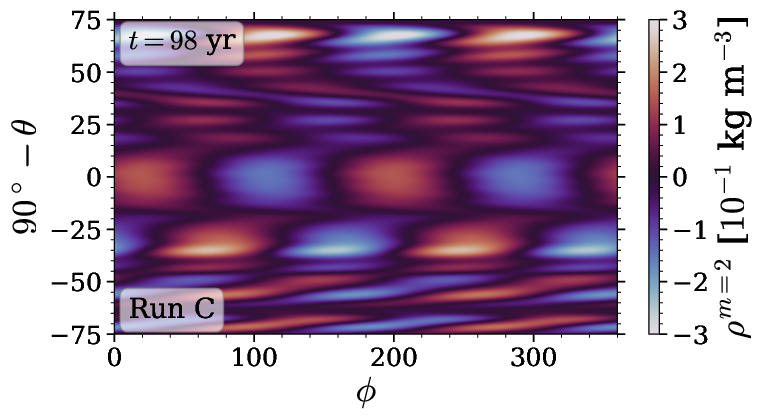}
\caption{Second nonaxisymmetric mode of density at $r=0.98R$ for each Run.}
\label{fig:rhom2}
\end{figure*}

\end{document}